\documentclass{sig-alternate}
\usepackage{color}
\usepackage{caption}
\usepackage{subcaption}
\usepackage{lscape}
\usepackage{cite}
\usepackage[lined,boxruled,linesnumbered]{algorithm2e}
\usepackage{epsfig}
\usepackage{enumitem}
\usepackage{amsmath}
\usepackage{amssymb}
\usepackage{hyperref}
\begin{document}
%%
%% --- Author Metadata here ---
%\conferenceinfo{WOODSTOCK}{'97 El Paso, Texas USA}
%%\CopyrightYear{2007} % Allows default copyright year (20XX) to be over-ridden - IF NEED BE.
%%\crdata{0-12345-67-8/90/01}  % Allows default copyright data (0-89791-88-6/97/05) to be over-ridden - IF NEED BE.
%% --- End of Author Metadata ---
%
\title{Human Social Interaction Modeling Using\\ Temporal Deep Networks}
\numberofauthors{6} %  in this sample file, there are a *total*
%%% of EIGHT authors. SIX appear on the 'first-page' (for formatting
%%% reasons) and the remaining two appear in the \additionalauthors section.
%%%
\author{
% 1st. author
\alignauthor 
Mohamed R. Amer\\
       \affaddr{SRI International}\\
       \email{Mohamed.Amer@sri.com}
% 2nd. author
\alignauthor 
Behjat Siddiquie\\
       \affaddr{SRI International}\\
       \email{Behjat.Siddiquie@sri.com}
% 3rd. author
\alignauthor  
Amir Tamrakar\\
       \affaddr{SRI International}\\
       \email{Amir.Tamrakar@sri.com}
\and
% 4th. author
\alignauthor 
David A. Salter\\
      \affaddr{SRI International}\\
      \email{David.Salter@sri.com}
% 5th. author
\alignauthor 
Brian J. Lande\\
      \affaddr{UC Santa Cruz}\\
      \email{brianlande@soe.ucsc.edu}
% 6th. author
\alignauthor 
Darius Mehri\\
 	  \affaddr{UC Berkeley}\\
 	  \email{darius\_mehri@berkeley.edu}
\and
% 7th. author
\alignauthor 
Ajay Divakaran\\
 	 \affaddr{SRI International}\\
	 \email{Ajay.Divakaran@sri.com}
}
\date{9 April 2015}
\maketitle
\begin{abstract}
We present a novel approach to computational modeling of social interactions based on modeling of essential social interaction predicates (ESIPs) such as joint attention and entrainment. Based on sound social psychological theory and methodology, we collect a new ``Tower Game'' dataset consisting of audio-visual capture of dyadic interactions labeled with the ESIPs. We expect this dataset to provide a new avenue for research in computational social interaction modeling. We propose a novel joint Discriminative Conditional Restricted Boltzmann Machine (DCRBM) model that combines a discriminative component with the generative power of CRBMs. Such a combination enables us to uncover actionable constituents of the ESIPs in two steps. First, we train the DCRBM model on the labeled data and get accurate (76\%-49\% across various ESIPs) detection of the predicates. Second, we exploit the generative capability of DCRBMs to activate the trained model so as to generate the lower-level data corresponding to the specific ESIP that closely matches the actual training data (with mean square error 0.01-0.1 for generating 100 frames). We are thus able to decompose the ESIPs into their constituent actionable behaviors. Such a purely computational determination of how to establish an ESIP such as engagement is unprecedented.
\end{abstract}
%
% A category with the (minimum) three required fields %A category including the fourth, optional field follows...
\section{Introduction} \label{sec:Intro}
\begin{figure*}[t]
\centering
\includegraphics[width=\textwidth]{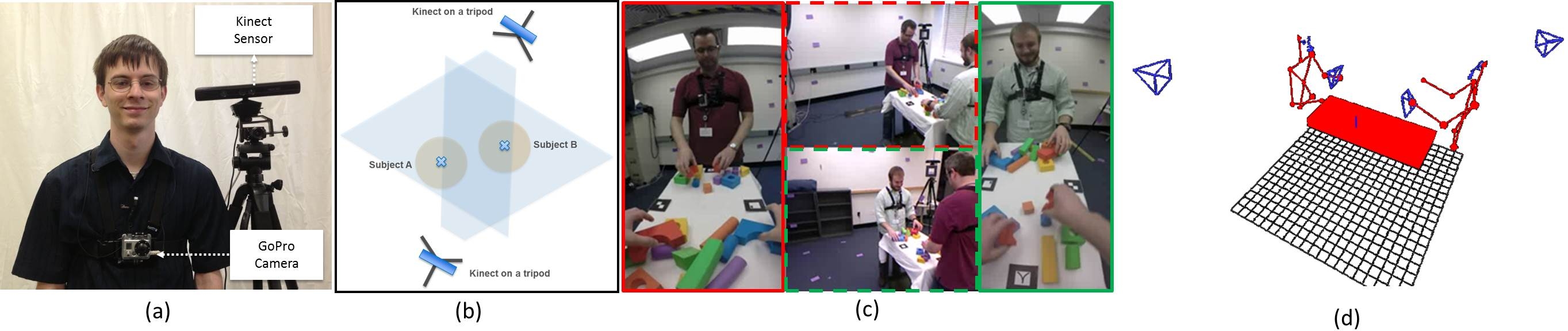}
\caption{(a) Our capture setup which includes a GoPro camera mounted on each participant's chest and a Kinect mounted on a tripod. (b) An overhead view of our capture setup involving the two participants. (c) Sample Data Collected: The image outlined in solid red shows the image captured from the GoPro camera mounted on player A (green shirt), while the image outlined in dashed red shows the image captured from the Kinect behind player A and is used to track the upper body of player B (red shirt). Similarly the image outlined in solid green is the image captured from the GoPro mounted on player B and the image outlined in dashed green is the image captured from the Kinect behind player B. (d) A view of our collected data projected in a unified coordinate framework.}
\label{fig:data_capture}
\end{figure*}
This research brings together multiple disciplines to explore the problem of social interaction modeling. The goal of this work is to leverage research in social psychology, computer vision, signal processing, and machine learning to better understand human social interactions.

As an example application, consider aid-workers or medical personnel deployed in a foreign country. During the course of their deployment, these workers often have to interact with people with whom they share little in common in terms of language, customs and culture. Reducing friction as well as increasing engagement between the workers and the populations they encounter can have an important bearing on the success of their mission. Therefore the ability to impart such professionals, with a general cross-cultural competency which would enable them to smoothly interact with the foreign populations they encounter would be extremely useful. With such an application in mind, we focus on identifying and automatically detecting predicates that facilitate social interactions irrespective of the cultural context. Since our interests lie in aspects of social  interactions that reduce conflict and build trust, we focus on social predicates that support rapport: joint attention, temporal synchrony, mimicry, and coordination.

Our orientation to social sensing departs significantly from existing methods~\cite{Baron_MIT1997, Picard_MIT1995} that focus on inferring internal or hidden mental states. Instead, inspired by a growing body of research~\cite{Tomasello_Harvard2001, Levinson_FARS2006, Louwerse_CS2012}, we focus on the process of social interaction. This research argues that social interaction is more than the meeting of two minds, with an additional emphasis on the cognitive, perceptual and motor explanations of the joint and coordinated actions that occur as part of these interactions~\cite{DiPaolo_FHN2012}. Our approach is guided by two key insights. The first is that apart from inferring the mental state of the other, social interactions require individuals to attend each other's movements, utterances and context to coordinate actions jointly with each other \cite{Sebanz_TCS2009}. The second insight is that social interactions involve reciprocal acts, joint behaviors along with nested events (e.g. speech, eye gaze, gestures) at various timescales and therefore demand adaptive and cooperative behaviors of their participants~\cite{Fantasia_FP2014}. 

Using the work of ~\cite{Jaegher_TCS2010} as a starting point, which emphasizes the interactive and cooperative aspects of the social interactions, we focus on detecting rhythmic coupling (also known as entrainment and attunement), mimicry (behavioral matching), movement simultaneity, kinematic turn taking patterns, and other measurable features of engaged social interaction. We established that behaviors such as \textit{joint attention} and \textit{entrainment} were the essential predicates of social interaction (ESIPs). With this in mind we focus on developing computational models of social interaction, that utilize multimodal sensing and temporal deep learning models to detect and recognize these ESIPs as well as discover their actionable constituents.

%Computer Vision and Machine Learning algorithms
Over the past decade, the fields of computer vision and machine learning have made significant advances. Furthermore, with the availability of complex sensors like Kinect, researchers are able to accurately track full human body poses~\cite{Shotton_CVPR2011}. This allowed for many different applications in such as activity recognition~\cite{Raptis_SCA2011}, facial feature tracking~\cite{Fanelli_IJCV2012}, and multimodal event detection~\cite{Koppula_ICML2013}. 

The sophistication of our problem requires a machine learning algorithm capable of jointly recognizing, correlating features, and generating multimodal data of dyadic social interactions.  Discriminative models focus on maximizing the separation between classes, however, they are often uninterpretable. On the other hand, generative models focus solely on modeling distributions and are often unable to incorporate higher level knowledge. Hybrid models tend to address these problems by combining the advantages of discriminative and generative models. They encode higher level knowledge as well as model the distribution from a discriminative perspective. We propose a novel hybrid model that allows us to recognize classes, correlate features, and generate social interaction data.

%Summary
This paper proposes new approach to machine learning that answers questions posed by social psychology. Our approach to social sensing is multimodal and attempts to detect the existence of features of social interaction, social interaction itself, and the qualitative and dynamic features of social interaction. We took a multimodal approach because humans must solve a variety of binding problems to effectively coordinate action. Coordination must span everything from postural sways, eye gazes, head pose, gestures, lexical choice, verbal pitch and intonation, etc.

\textit{Our contributions} are 3-fold:
\begin{itemize}[itemsep=-1pt,topsep=-2pt, partopsep=1pt]
\item A new problem of computational modeling of essential social interaction predicates (ESIPs). Starting from a socio-psychological framework, we demonstrate the use of multimodal sensors and temporal deep learning models to uncover actionable constituents of ESIPs. 
\item A new dataset, \textit{Tower Game Dataset}, for analyzing social interaction predicates. The dataset consists of multimodal recordings of two players participating in a tower building game, in the process communicating and collaborating with each other. The dataset has been annotated with ESIPs and will be made publicly available. We believe that it will foster new research in the area of computational social interaction modeling.
\item A novel model, Discriminative Conditional Restricted Boltzmann Machine (DCRBM), that introduces a discriminative component to Conditional Restricted Boltzmann Machines (CRBM). The discriminative component enables DCRBMs to directly learn classification models while retaining all the advantages of CRBMs, including their ability to generate missing data. Results on the \textit{Tower Game Dataset} demonstrate that DCRBMs can effectively detect ESIPs as well decompose ESIPs into their constituent actionable behaviors.\\ 
\end{itemize}
\textit{Paper organization}: In sec.~\ref{sec:related_work} we discuss prior work. In sec.~\ref{sec:Approach} we specify our model, then we explain inference and learning. In sec.~\ref{sec:experiments} we describe our dataset and demonstrate the quantitative results of our approach. In sec.~\ref{sec:conclusion} we conclude.
\section{Related Work}
\label{sec:related_work}
\noindent{\bf Social Psychology:} The study of social interactions and their associated sociological and psychological implications has received a lot of attention from social science researchers~\cite{Levinson_FARS2006, DiPaolo_FHN2012, Bernieri_JNVB1988}. Early research focused on the ``Theory of Mind" according to which individuals ascribe mental states to themselves and others~\cite{Baron_MIT1997}, a line of thinking that largely inspired much of the initial work on affective computing. However, more recent work has shown that apart from inferring each other's mental states, an important challenge for participants of a social interaction is to pragmatically sustain sequences of action where the action is tightly coupled to one another via multiple channels of observable information (e.g. visible kinematic information, audible speech). In other words, social interactions require dynamically coupled interpersonal motor coordination from their participants~\cite{Sebanz_TCS2009}. Moreover, detecting coupled behaviors such as kinematic turn taking or simultaneity in movements can help in recognizing engaged social interactions~\cite{Jaegher_TCS2010}. \\
\noindent{\bf Affective Computing:} refers to the study and development of systems that can automatically detect human affect~\cite{Picard_MIT1995, Calvo_TAC2010}. Affective computing has long been an active research area due to its utility in a variety of applications that require realistic Human Computer Interaction, such as online tutoring~\cite{DMello_IS2007} and health screenings~\cite{Ghosh_ICMI2014}. The goal here is to detect the overall mental or emotional state of the person based on external cues. This is typically done based on speech~\cite{Amer_ICASSP2014}, facial expressions~\cite{Tian_PAMI2001}, gesture/posture~\cite{Mota_CVPRW2003} and multimodal cues~\cite{Siddiquie_ICME2013, Ramirez_ACII2011, Amer_WACV2014}. There has also been work on modeling activities and interactions involving multiple people~\cite{Ryoo_IJCV2009,Lan_PAMI2012,Rehg_CVPR2013}. However, most of this work deals with short duration task-oriented activities~\cite{Ryoo_IJCV2009,Lan_PAMI2012} with a focus on their physical aspects. There has been a recent interest in modeling interactions with a focus on the rich and complex social behaviors that they elicit along with their affective impact on the participants ~\cite{Rehg_CVPR2013}.  \\
\noindent{\bf Hybrid Models:} consist of a generative model, which usually learns a feature representation of low level input, and a discriminative model for higher level reasoning. Recent work has empirically shown that generative models which learn a rich feature representation tend to outperform discriminative models that rely solely on hand-crafted features \cite{Perina_PAMI2012}. Hybrid models can be divided into three groups,  joint methods \cite{Lasserre_CVPR2006, Mccallum_AAAI2006, Larochelle_ICML2008, Druck_ICML2010}, iterative methods\cite{Sminchisescu_CVPR2006, Fujino_PAMI2008}, and staged methods \cite{Jebara_MLR2004, Bosch_TPAMI2008, Li_CVPR2011, Ranzato_CVPR2011, Perina_PAMI2012}. Joint methods optimize a single objective function which consists of both the generative and discriminative energies. Iterative methods consist of a generative and a discriminative model that are trained in an iterative manner, influencing each other. In staged methods, both models are trained separately, with the discriminative model being trained on representations learned by the generative model. Classification is performed after projecting the samples into a fixed-dimensional space induced by the generative model. \\
\noindent{\bf Deep Networks:} are able to learn rich features in an unsupervised manner, this is what makes deep learning very powerful.  They have been successfully applied to many problems \cite{Bengio_FTML2009}.  Restricted Boltzmann Machines (RBMs) form the building blocks in deep networks models \cite{Hinton_NC2006, Salakhutdinov_Science2006}. In \cite{Hinton_NC2006, Salakhutdinov_Science2006}, the networks are trained using the Contrastive Divergence (CD) algorithm \cite{Hinton_NC2002}, which demonstrated the ability of deep networks to capture the distributions over the features efficiently and to learn complex representations.  RBMs can be stacked together to form deeper networks known as Deep Boltzmann Machines (DBMs), which capture more complex representations. Recently, deep networks based temporal models, capable of modeling a more temporally rich set of problems have been proposed. These include Conditional RBMs (CRBMs) \cite{Taylor_JMLR2011} and Temporal RBMs (TRBMs) \cite{Sutskever_AISTATS2007, Sutskever_NIPS2008, Hausler_CoRR2012}. CRBMs have been successfully used in both visual and audio domains. They have been used for modeling human motion \cite{Taylor_JMLR2011}, tracking 3D human pose \cite{Taylor_CVPR2010} and phone recognition \cite{Mohamed_ICML2009}. TRBMs have been applied for transferring 2D and 3D point clouds \cite{Zeiler_NIPS2011}, transition based dependency parsing \cite{Garg_ACL2011}, and polyphonic music generation \cite{Lewandowski_ICML2009}.\\
\section{Approach}\label{sec:Approach}
In this section we describe our approach.  We first review similar prior work, next we define our model, formulate its inference, and finally show how the model parameters are learned.
\subsection{Review of Prior Models}\label{sec:Model}
\begin{figure}
\centering
\begin{subfigure}[h]{0.4\textwidth}
\includegraphics[width=\textwidth]{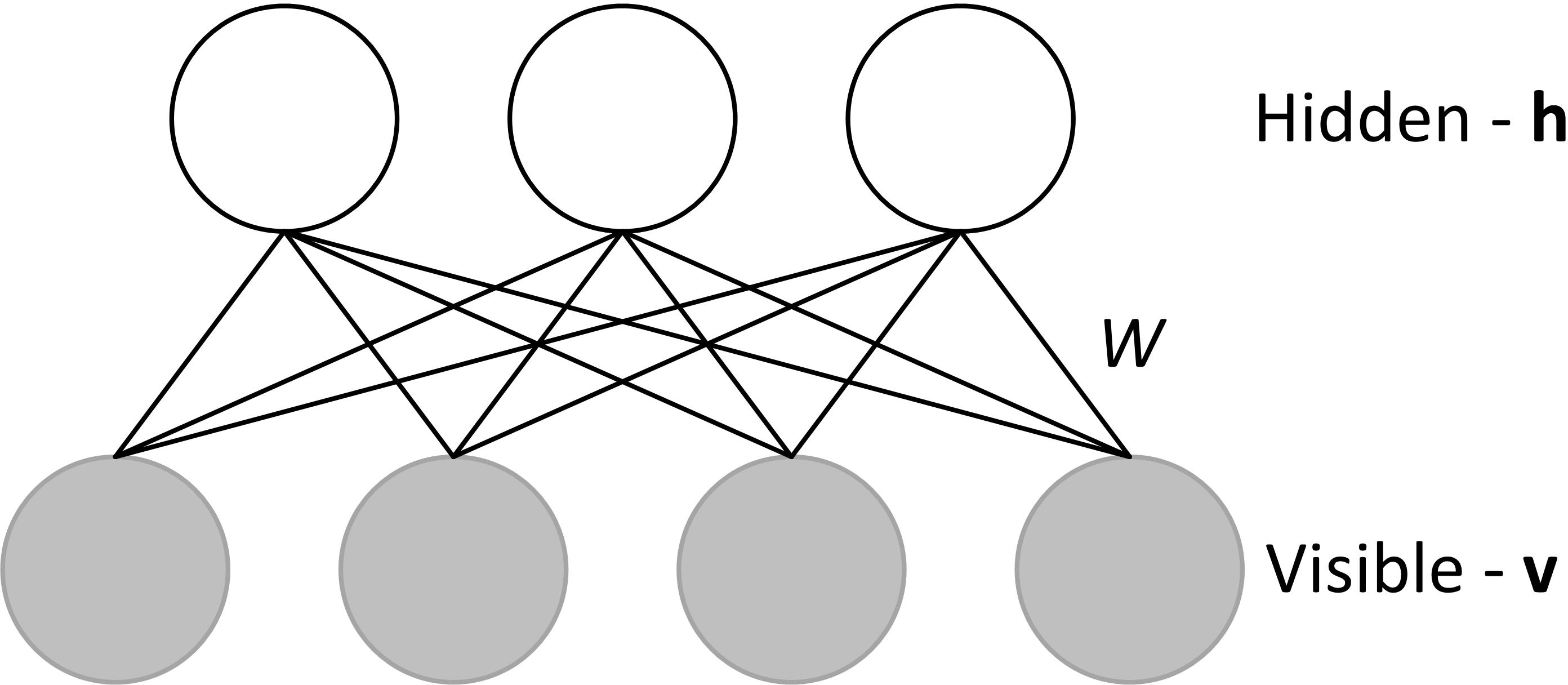}
\caption{RBM}
\label{fig:RBM}
\end{subfigure}
\vfill%
\begin{subfigure}[h]{0.4\textwidth}
\includegraphics[width=\textwidth]{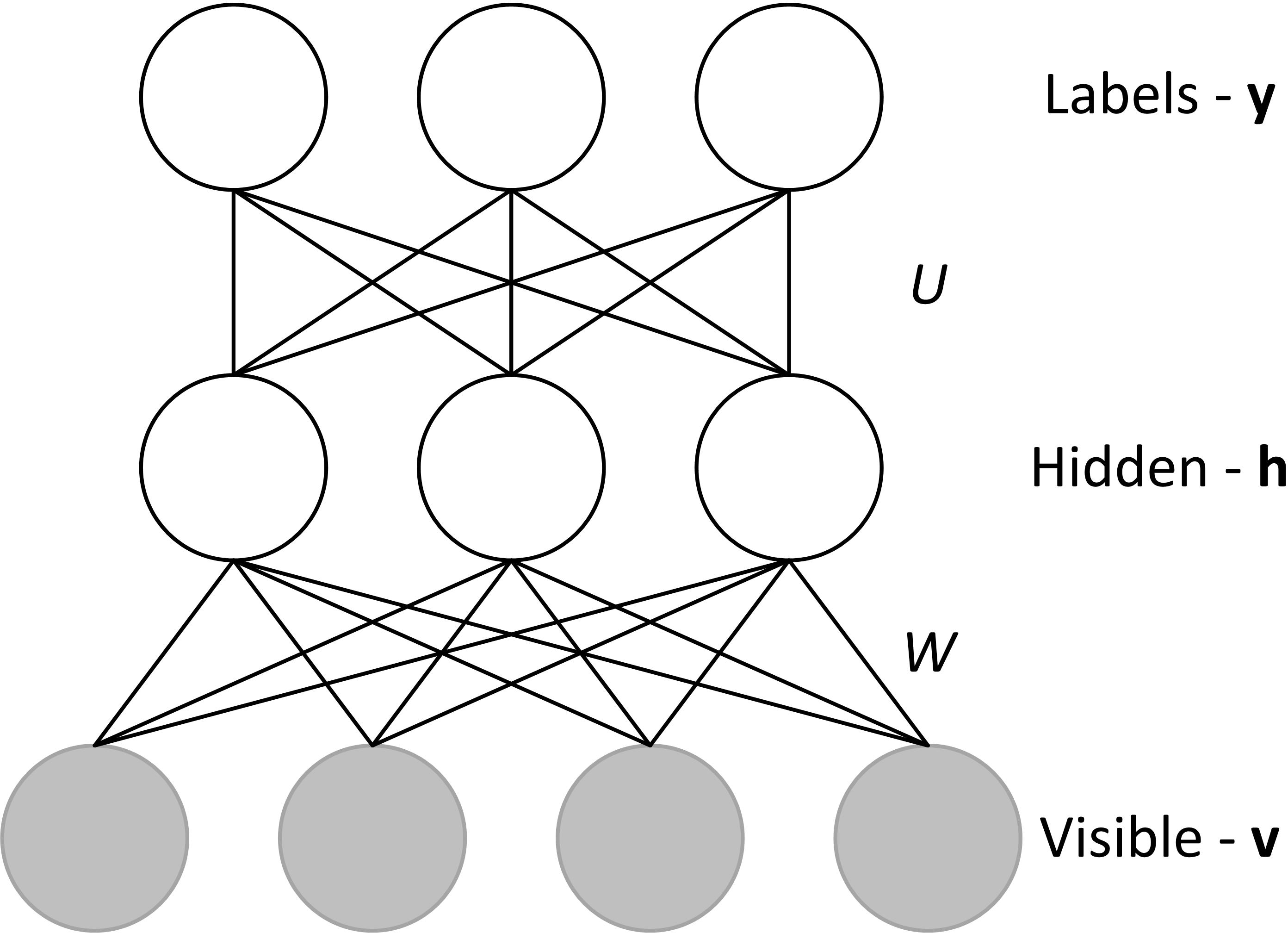}
\caption{DRBM}
\label{fig:DRBM}
\end{subfigure}
\vfill%
\begin{subfigure}[h]{0.4\textwidth}
\includegraphics[width=\textwidth]{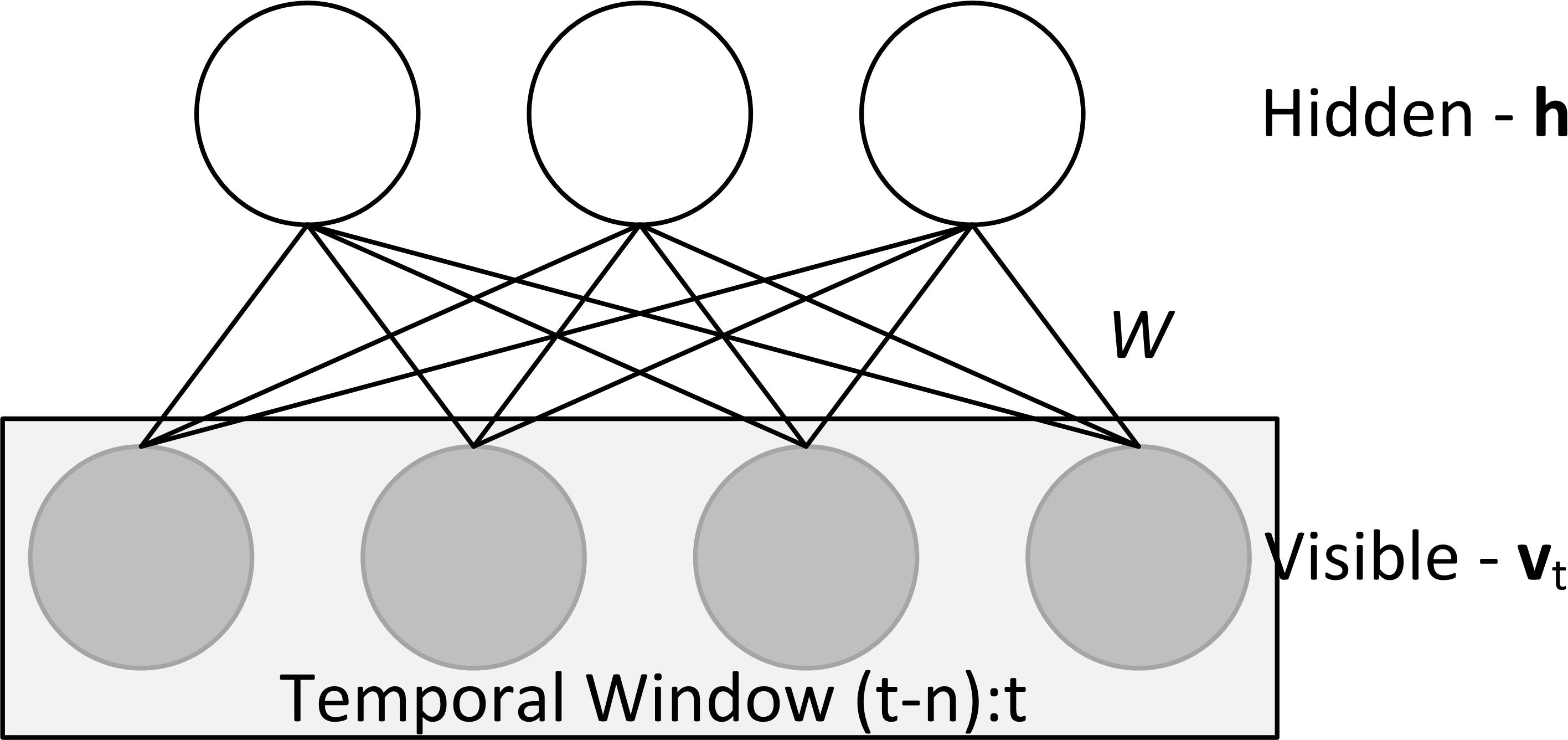}
\caption{CRBM}
\label{fig:CRBM}
\end{subfigure}
\vfill%
\begin{subfigure}[h]{0.4\textwidth}
\includegraphics[width=\textwidth]{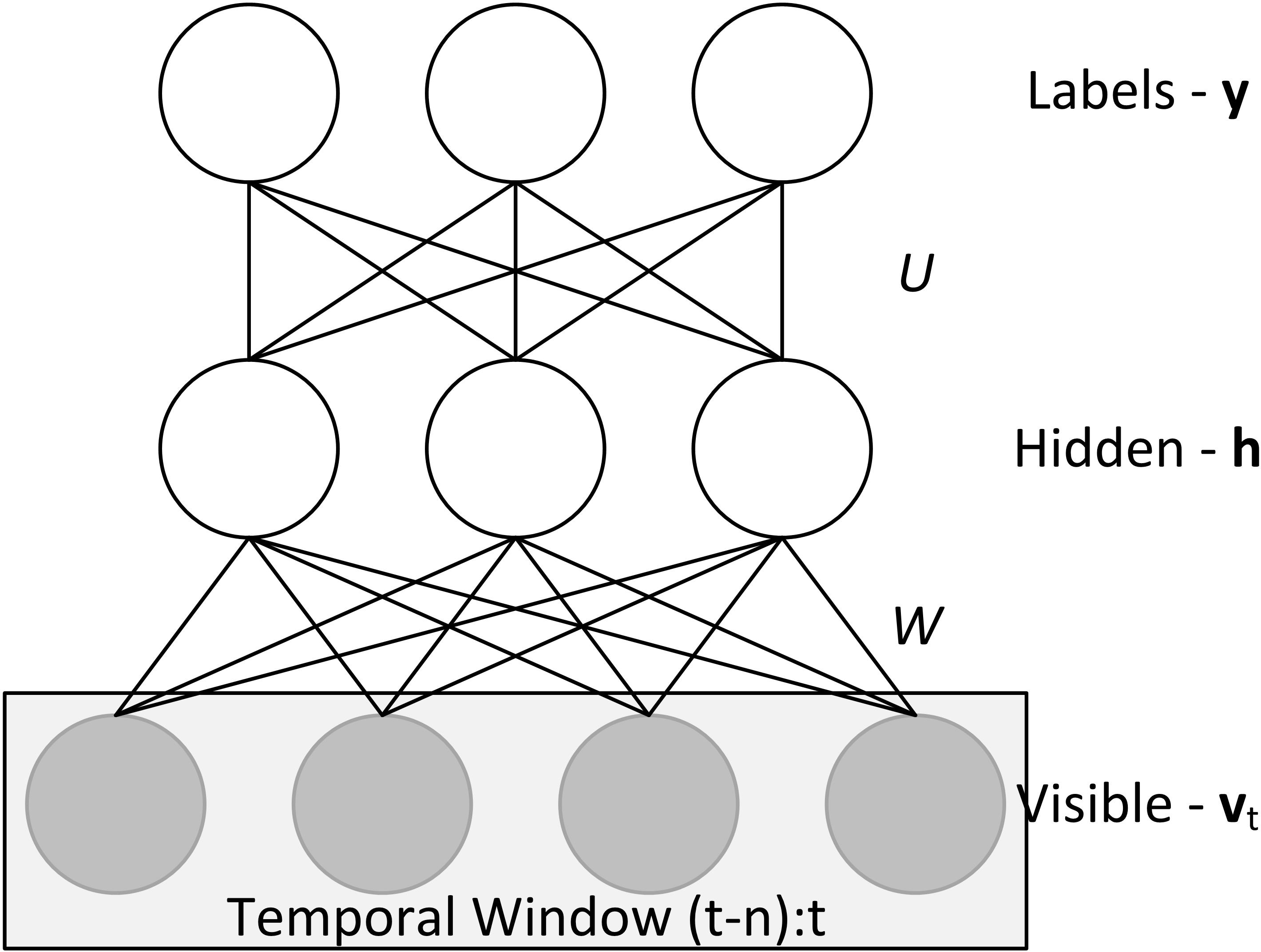}
\caption{DCRBM}
\label{fig:DCRBM}
\end{subfigure}
\caption{Deep learning models described in sections~\ref{sec:Approach}. (a) RBM and (c) CRBM are generative models. (b)DRBM and (d)DCRBM are discriminatively trained hybrid models.}
\label{fig:ModelFigure}
\end{figure}
\noindent{\bf Restricted Boltzmann Machines} \cite{Salakhutdinov_Science2006}: An RBM (Fig.~\ref{fig:RBM}) defines a probability distribution $p_{\text{R}}$ as a Gibbs distribution (\ref{eqn:RBM}), where ${\bf v}$ is a vector of visible nodes, ${\bf h}$ is a vector of hidden nodes. $E_{\text{R}}$ is the energy function and $Z$ is the partition function which ensures that the distribution is valid. The parameters ${\boldsymbol \theta}_{\text{R}}$ to be learned are ${\bf a}$ and ${\bf b}$ the biases for ${\bf v}$ and ${\bf h}$ respectively and the weights ${\it W}$. The RBM architecture is defined as fully connected between layers, with no lateral connections.  This architecture implies that {\bf v} and {\bf h} are factorial given one of the two vectors.  This allows for the exact computation of $p({\bf v}|{\bf h})$ and $p({\bf h}|{\bf v})$.
\begin{equation}
\begin{array}{rcl}
p_{\text{R}}({\bf v},{\bf h};{\boldsymbol \theta}_{\text{R}})&=&\exp[-E_{\text{R}}({\bf v},{\bf h})]/Z({\boldsymbol \theta}_{\text{R}}),\\
\\
Z({\boldsymbol \theta}_{\text{R}})&=&\sum_{{\bf v},{\bf h}}\exp[-E_{\text{R}}({\bf v},{\bf h})],\\
\\
{\boldsymbol \theta}_{\text{R}}&=&\{{\bf a},{\bf b},{\it W}\}
\end{array}
\label{eqn:RBM}
\end{equation}
In case of binary valued data $v_i$ is defined as  a logistic function. In case of real valued data, $v_i$ is defined as a multivariate Gaussian distribution with a unit covariance. A binary valued hidden layer $h_j$ is defined as a logistic function\footnote{The logistic function $\sigma(\cdot)$ for a variable $x$ is defined as $\sigma(x)=(1+exp(-x))^{-1}$.}. This is done because we want the hidden layer to be a sparse binary code (empirically proven to be better \cite{Taylor_JMLR2011, Sutskever_AISTATS2007}). (\ref{eqn:PRBM}) shows the probability distributions for $v$
\begin{equation}
\begin{array}{rcll}
p(v_{i} = 1 |{\bf h})&=&\sigma(a_{i}+\sum_{j} h_{j} w_{ij}),\quad& \text{Binary,}\\
\\
p(v_{i}|{\bf h})&=&\mathcal{N}(a_{i}+\sum_{j} h_{j}w_{ij},1),\quad& \text{Real,}\\
\\
p(h_{j} = 1 |{\bf v})&=&\sigma(b_{j}+\sum_{i} v_{i} w_{ij}),\quad& \text{Binary.}
\end{array}
\label{eqn:PRBM}
\end{equation} 
The energy function $E_{\text{R}}$ for binary $v_i$ is defined as in (\ref{eqn:EBRBM}).
\begin{equation}
E_{\text{R}}({\bf v},{\bf h})=-\sum_{i} a_{i} v_{i}- \sum_{j} b_{j} h_{j}- \sum_{i,j} v_{i}w_{i,j} h_{j},
\label{eqn:EBRBM}
\end{equation}
while, the energy function $E_{\text{R}}$ is slightly modified  to allow for the real valued ${\bf v}$ as shown in (\ref{eqn:ERRBM}).
\begin{equation}
E_{\text{R}}({\bf v},{\bf h})=-\sum_{i} \frac{(a_{i}-v_{i})^2}{2} - \sum_{j} b_{j} h_{j}- \sum_{i,j} v_{i}w_{i,j} h_{j}
\label{eqn:ERRBM}
\end{equation}

\noindent{\bf Discriminative Restricted Boltzmann Machines} \cite{Larochelle_ICML2008}: DRBMs are a natural extension of RBMs which have an additional discriminative term for classification. They are based on the model in \cite{Larochelle_ICML2008}. DRBM (Fig.~\ref{fig:DRBM}) defines a probability distribution $p_{\text{D}}$ as a Gibbs distribution (\ref{eqn:DRBM}). 
\begin{equation}
\begin{array}{rcl}
p_{\text{DR}}({\bf y},{\bf v},{\bf h}|{\bf v};{\boldsymbol \theta}_{\text{DR}})&=&\exp[-E_{\text{DR}}( {\bf y},{\bf v},{\bf h})]/Z({\boldsymbol \theta}_{\text{DR}}),\\
\\
Z({\boldsymbol \theta}_{\text{DR}})&=&\sum_{{\bf y},{\bf v},{\bf h}}\exp[-E_{\text{DR}}({\bf y},{\bf v},{\bf h})]\\
\\
{\boldsymbol \theta}_{\text{C}}&=&\{{\bf a},{\bf b},{\bf s},{\it W},{\it U}\}
\end{array}
\label{eqn:DRBM}
\end{equation}
The probability distribution over the visible layer will follow the same distributions as in (\ref{eqn:PRBM}). The hidden layer ${\bf h}$ is defined as a function of the labels $y$ and the visible nodes ${\bf v}$. Also, a new probability distribution for the classifier is defined to relate the label $y$ to the hidden nodes ${\bf h}$ as in (\ref{eqn:PDRBM}).

\begin{equation}
\begin{array}{rcl}
p(v_{i}|{\bf h})&=&\mathcal{N}(a_{i}+\sum_{j} h_{j}w_{ij},1),\\
\\
p(h_{j} = 1 |y_k,{\bf v})&=&\sigma(b_{j}+ u_{j,k}+\sum_{i} v_{i} w_{ij}),\\
\\
p(y_k|{\bf h})&=&\frac{e^{s_k+\sum_j u_{j,k}h_j}}{\sum_{k^*} e^{s_{k^*}+\sum_j u_{j,k^*}h_j}}
\end{array}
\label{eqn:PDRBM}
\end{equation}
The new energy function $E_{\text{DR}}$ is defined similar to (\ref{eqn:EDRBM}),
\begin{eqnarray}
\begin{array}{rcl}
E_{\text{D}}({\bf y},{\bf v},{\bf h})&=&-\sum_{i} (a_{i}-v_{i})^2/2 - \sum_{j} b_{j} h_{j} - \sum_{k} s_{k} y_{k}\\
\\
&& -\sum_{i,j} v_{i}w_{i,j} h_{j} - \sum_{j,k} h_{j}u_{j} y_{k}
\end{array}
\label{eqn:EDRBM}
\end{eqnarray}

\noindent{\bf Conditional Restricted Boltzmann Machines} \cite{Taylor_JMLR2011}: CRBMs are a natural extension of RBMs for modeling short term temporal dependencies. A CRBM (Fig.~\ref{fig:CRBM}) is an RBM which takes into account history from the previous time instances $[(t-N),\hdots,(t-1)]$ at time $(t)$. This is done by treating the previous time instances as additional inputs. Doing so does not complicate inference\footnote{Some approximations have been made to facilitate efficient training and inference, more details are available in \cite{Taylor_JMLR2011}.}. A CRBM defines a probability distribution $p_{\text{C}}$ as a Gibbs distribution (\ref{eqn:CRBM}). 
\begin{equation}
\begin{array}{rcl}
p_{\text{C}}({\bf v}_{t},{\bf h}_{t}|{\bf v}_{<t};{\boldsymbol \theta}_{\text{C}})&=&\exp[-E_{\text{C}}({\bf v}_{t},{\bf h}_{t}|{\bf v}_{<t})]/Z({\boldsymbol \theta}_{\text{C}}),\\
\\
Z({\boldsymbol \theta}_{\text{C}})&=&\sum_{{\bf v},{\bf h}}\exp[-E_{\text{C}}({\bf v}_{t},{\bf h}_{t}|{\bf v}_{<t})]\\
\\
{\boldsymbol \theta}_{\text{C}}&=&\{{\bf a},{\bf b},{\it A},{\it B},{\it W}\}
\end{array}
\label{eqn:CRBM}
\end{equation}
The additional inputs from previous time instances are modeled as directed autoregressive edges from the past $N$ visible nodes and the past $M$ hidden layers, where, $N$ does not have to be equal to $M$.  The concatenated history vector is defined as ${\bf v}_{<t}$.  The probability distributions are defined in (\ref{eqn:PCRBM}).
\begin{equation}
\begin{array}{rcl}
p(v_{i}|{\bf h},{\bf v}_{<t})&=&\mathcal{N}(a_{i} + \sum_{n}A_{n,i} v_{n,<t} + \sum_{j} h_{j}w_{ij},1),\\
\\
p(h_{j} = 1 |{\bf v},{\bf v}_{<t})&=&\sigma(b_{j} + \sum_{m}B_{m,j} v_{m,<t} + \sum_{i} v_{i} w_{ij}).
\end{array}
\label{eqn:PCRBM}
\end{equation}
The new energy function $E_{\text{C}}({\bf v}_{t},{\bf h}_{t}|{\bf v}_{<t})$ in (\ref{eqn:ECRBM}) is defined in a manner similar to that of the RBM (\ref{eqn:ERRBM}).
\begin{equation}
\begin{array}{rcl}
E_{\text{C}}({\bf v}_{t},{\bf h}_{t}|{\bf v}_{<t})&=&-\sum_{i} \frac{(c_{i}-v_{i,t})^2}{2} - \sum_{j} d_{j} h_{j,t}\\
\\
&&- \sum_{i,j} v_{i,t} w_{i,j} h_{j,t},\\
\end{array}
\label{eqn:ECRBM}
\end{equation}
where,
\begin{equation}
 c_{i}= a_{i} + \sum_{n}A_{n,i} v_{n,<t}, \quad d_{j}=b_{j} + \sum_{m}B_{m,j} v_{m,<t}.\nonumber
\label{eqn:CDCRBM}
\end{equation}
Note that $A$ and $B$ are matrices of concatenated vectors of previous time instances of ${\bf a}$ and ${\bf b}$.
\subsection{Model}
\noindent{\bf Discriminative Conditional Restricted Boltzmann Machines}: (DCRBMs) are a natural extension of CRBMs which have an additional discriminative term for classification. They are based on the model in \cite{Larochelle_ICML2008}, generalized to account for temporal phenomenon using CRBMs. DCRBMs (Fig.~\ref{fig:DCRBM}) are a simpler version of the Factored Conditional Restricted Boltzmann Machines \cite{Taylor_JMLR2011} and Gated Restricted Boltzmann Machines \cite{Memisevic_CVPR2007}. Both these models incorporate labels in learning representations, however, they use a more complicated potential which involves three way connections into factors. DCRBM defines a probability distribution $p_{\text{DC}}$ as a Gibbs distribution (\ref{eqn:DCRBM}). 
\begin{equation}
\begin{array}{c}
p_{\text{DC}}({\bf y}_{t},{\bf v}_{t},{\bf h}_{t}|{\bf v}_{<t};{\boldsymbol \theta}_{\text{DC}})=\exp[-E_{\text{DC}}( {\bf y}_{t},{\bf v}_{t},{\bf h}_{t}|{\bf v}_{<t})]/Z({\boldsymbol \theta}_{\text{DC}}),\\
\\
Z({\boldsymbol \theta}_{\text{DC}})=\sum_{{\bf y}_{t},{\bf v}_{t},{\bf h}_{t}}\exp[-E_{\text{DC}}({\bf y}_{t},{\bf v}_{t},{\bf h}_{t}|{\bf v}_{<t})],\\
\\
{\boldsymbol \theta}_{\text{DC}}=\{{\bf a},{\bf b},{\bf s},{\it A},{\it B},{\it W},{\it U}\}.
\end{array}
\label{eqn:DCRBM}
\end{equation}
The probability distribution over the visible layer will follow the same distributions as in (\ref{eqn:PDRBM}). The hidden layer ${\bf h}$ is defined as a function of the labels $y$ and the visible nodes ${\bf v}$. A new probability distribution for the classifier is defined to relate the label $y$ to the hidden nodes ${\bf h}$ is defined as in (\ref{eqn:PDCRBM}).

\begin{equation}
\begin{array}{l}
p(v_{i,t}|{\bf h}_{t},{\bf v}_{<t})=\mathcal{N}(a_{i} + \sum_{n}A_{n,i} v_{n,<t} + \sum_{j} h_{j}w_{ij},1),\\
\\
p(h_{j,t} = 1 |y_{t},{\bf v}_{t},{\bf v}_{<t})=\\
\quad\quad\quad\quad\quad \sigma(b_{j} + u_{j,k} + \sum_{i} v_{i,t} w_{ij} + \sum_{m}B_{m,j} v_{m,<t}),\\
\\
p(y_{k,t}|{\bf h})=\frac{e^{s_k+\sum_j u_{j,y}h_j}}{\sum_{k^*} e^{s_{k^*}+\sum_j u_{j,k^*}h_j}}.
\end{array}
\label{eqn:PDCRBM}
\end{equation}
The new energy function $E_{\text{DC}}$ is defined similar to that of the DRBM (\ref{eqn:EDRBM}).
\begin{equation}
\begin{array}{c}
E_{\text{DC}}({\bf y}_{t},{\bf v}_{t},{\bf h}_{t}|{\bf v}_{<t})=-\sum_{i} (c_{i}-v_{i,t})^2/2 - \sum_{j} d_{j,k} h_{j,t} \\
\\
- \sum_{k} s_{k} y_{k,t} - \sum_{i,j} v_{i,t} w_{i,j} h_{j,t} - \sum_{j,k} h_{j,t} u_{j,k} y_{k,t}
\end{array}
\label{eqn:EDCRBM}
\end{equation}
where,
\begin{equation}
c_{i}= a_{i} + \sum_{n}A_{n,i} v_{n,<t},\quad d_{j,k}=b_{j,t} + u_{j,k} + \sum_{m}B_{m,j} v_{m,<t}.  \nonumber
\label{eqn:CDEDCRBM}
\end{equation}

Note that $A$ and $B$ are matrices of concatenated vectors of previous time instances of ${\bf a}$ and ${\bf b}$.

\subsection{Inference and Learning}\label{sec:InferenceLearning}
\noindent{\bf Inference:} For classification we use a bottom up approach, where we maximize the posterior distribution, $p_{\text{DC}}(y_{t,k}|{\bf v}_{t},{\bf v}_{<t})$, over all the labels. This is equivalent to activating the hidden layer given the visible layer ${\bf v}_{t}$, visible layer history ${\bf v}_{<t})$, and label $y_{t,k}$ as shown in (\ref{eqn:DCRBMInference}).
\begin{equation}
\begin{array}{c}
y_t = \arg \max_{k}\quad p_{\text{DC}}(y_{t,k}|{\bf v}_{t},{\bf v}_{<t}),\quad \text{where,}\\
\\
p_{\text{DC}}(y_{t,k}|{\bf v}_{t},{\bf v}_{<t})=  \frac{e^{s_k}\prod_{j}\left(1+e^{s_k + d_{j,y} + \sum_i v_{i,t}w_{ij}} \right)}{\sum_{k^*} e^{s_{k^*}} \prod_{j} \left(1+e^{s_{k^*} + d_{j,{k^*}} + \sum_i v_{i,t}w_{ij}}\right)}.
\end{array}
\label{eqn:DCRBMInference}
\end{equation}
For generation we use a combination of top-down/bottom-up depending on the type of generation by activating the required layers given the available data, as in (\ref{eqn:PDCRBM}). Figures~\ref{fig:CrossModal1} and~\ref{fig:UniModal1} show the two cases. The first case (Fig.~\ref{fig:CrossModal1}) deals with partial missing data, where we have partial data for the hidden layer $v_{t}$ as well as the label $y$, and our goal is to generate the missing part of the $v_{t}$. The second case (Fig.~\ref{fig:UniModal1}) is when we have a fully missing visible layer $v_{t}$ and our goal is to generate it given only the class label $y$. For both cases we assume we have access to some history information.

\noindent{\bf Learning:} Learning our model is done using Contrastive Divergence (CD) \cite{Hinton_NC2002}. The update equations of the dynamically changing bases $\Delta{\bf c}$ and $\Delta{\bf d}$ are obtained by first updating $\Delta A$ and $\Delta B$ as in the case of the real valued CRBM (\ref{eqn:CRBM}) and then combining them with $\Delta a$ and $\Delta b$. 
\begin{equation}
\begin{array}{rclcl}
\Delta w_{i,j}&\propto&\langle v_{i}h_{j}\rangle _{data} &-& \langle v_{i}h_{j}\rangle _{recon},\\ 
\Delta u_{j,k}&\propto&\langle y_{k}h_{j}\rangle _{data} &-& \langle y_{k}h_{j}\rangle _{recon},\\ 
\Delta a_{i}&\propto&\langle v_{i}\rangle _{data} &-& \langle v_{i}\rangle _{recon},\\
\Delta b_{j}&\propto&\langle h_{j}\rangle _{data} &-& \langle h_{j}\rangle _{recon},\\
\Delta s_{k}&\propto&\langle y_{k}\rangle _{data} &-& \langle y_{k}\rangle _{recon},\\
\Delta A_{k,i,t-n}&\propto& v_{k,t-n}(\langle v_{i,t}\rangle _{data} &-& \langle v_{i,t}\rangle _{recon}),\\
\Delta B_{i,j,t-m}&\propto& v_{i,t-m}(\langle h_{j,t}\rangle _{data} &-& \langle h_{j,t}\rangle _{recon}),
\end{array}
\label{eqn:CD}
\end{equation}
where $\langle\cdot\rangle _{data}$ is the expectation with respect to the data distribution and $\langle\cdot\rangle _{recon}$ is the expectation with respect to the reconstructed data. The reconstruction is generated by first sampling $p(h_{j}=1|{\bf v},y)$ for all the hidden nodes in parallel. The visible nodes are then generated by sampling $p(v_{i}|{\bf h})$ for all the visible nodes in parallel. Finally, the label nodes are generated using $p(y|{\bf h})$ using (\ref{eqn:PDCRBM}).

\section{Experiments} \label{sec:experiments}

In this section, we first discuss existing activity recognition and affective computing datasets. Next we describe the collection and annotation of our \textit{Tower Game Dataset}, which contains recordings of two players building a tower and in the process engaging in a variety of interactive behaviors. Finally, we describe our experimental results on this dataset, demonstrating the effectiveness of our DCRBM model.

\subsection{Datasets}

Most existing activity recognition benchmarks -- e.g., the Weizmann, Trecvid, PETS04, CAVIAR, IXMAS, Hollywood datasets, Olympic Sports and UCF-100 --  contain relatively simple and repetitive actions involving a single person \cite{Chaquet_CVIU2013}.  On the other hand, group activity recognition datasets such as UCLA Courtyard, UT-Interactions, Collective Activity datasets, and Volleyball dataset, lack rich social dynamics.  

Other relevant datasets include the Multimodal Dyadic Behavior (MMDB) dataset~\cite{Rehg_CVPR2013}, which focuses on analyzing dyadic social interactions between adults and children in a developmental context. This dataset was collected in a semi-structured format, where children interact with an adult examiner in a series of pre-planned games. However, due to its narrow focus on analysis of social behaviors to diagnose developmental disorders in children, we believe it is not general enough. Another dataset is the Mimicry database~\cite{Sun_ACII2011} which focuses on studying social interactions between humans with the aim of analyzing mimicry in human-human interactions. This dataset was collected in an unstructured format where the two humans talk to each other about different subjects.

There are a number of issues with the aforementioned datasets, including: (a) unnatural, acted activities in constrained scenes; (b) limited spatial and temporal coverage; (c) poor diversity of activity classes; (d) Lack of rich social interactions; (e) Narrow focus on a single behavior (e.g.  mimicry); and (f) Unstructured or uncontrolled collection setup. Hence, we propose our new Tower Game Dataset to address the above issues.\\
\\
\noindent{\bf Tower Game Dataset} is a simple game of tower building often used in social psychology to elicit different kinds of interactive behaviors from the participants. It is typically played between two people working with a small fixed number of simple toy blocks that can be stacked to form various kinds of towers. We choose these tower games as they force the players to engage and communicate with each other in order to achieve the objectives of the game, thereby evoking behaviors such as \textit{joint-attention} and \textit{entrainment} from the participants. The game, due to its simplicity, allows for total control over the variables of an interaction. Due to the small number of blocks involved, the number of potential moves (actions) is limited. Also since the game involves interacting with physical objects, \textit{joint-attention} is mediated through concrete objects. Furthermore, only two players are involved, ensuring that we can stay in the realm of dyadic interactions.

There are many different variants of the game. We settled on two variants designed to elicit maximum communication between the players, namely, (i) the \textbf{architect-builder} variant and (ii) the \textbf{distinct-objective} variant. Furthermore, in order to maximize the amount of non-verbal communication, we prohibited the participants from verbally communicating with each other.

The \textbf{architect-builder} variant involves one participant playing the role of the architect, who decides the kind of tower to build and how to build it. The second participant is the builder, who has control of all the building blocks and is the only one actually manipulating the blocks. The goal here is for the architect to communicate to the builder how to build the tower so that builder can build the desired tower.

The \textbf{distinct-objective} variant is slightly more complicated and is designed to elicit more interaction between the players. In this variant, each player is given half of the building blocks required to build the tower. Each player is also given a particular rule, restricting the configuration of the tower being built, that they are required to enforce. An example rule could be that no two blocks of the same color may be placed such that they are touching each other. To make the play interesting, each player only knows their own rule and is not aware of rule given to the other player. The rules are selected at random from a small rule book. While certain combinations of rules may result in some conflict between the objectives of the two players, this is typically not the case. However, since each player needs to adhere to their rule, it means that they will need to correct an action taken by the other if it conflicts with their rule. In the process, each player also tries to figure out the rule assigned to the other player so that the process of building the tower is more efficient. Also, when the subjects played multiple sessions of this game, the pieces used were changed and the area of the table upon which they could place blocks was reduced in size.\\
\\
\noindent{\bf Capture Setup:} Our sensors include a pair of Kinect cameras that record color videos, depth video and track skeletons of the players and a pair of GoPro cameras mounted on the chest of each player (Fig.~\ref{fig:data_capture}(a)). External lapel microphones were attached to the GoPro cameras. However, the audio captured from them was used only for data synchronization purposes. Since the players were not allowed to verbally communicate with each other, very little speech (or paralinguistic) data exists. 

In order to ensure optimal data capture from the Kinect cameras (i.e. minimal occlusions and optimal skeleton tracking), they were mounted on tripods facing one another, slightly to the right and back of each of the participants and slightly elevated, ensuring that each camera got an unobstructed view of the other participant. The overhead layout is shown in Fig.~\ref{fig:data_capture}(b). These videos are of VGA resolutions (640x480) and were captured at 30Hz. The GoPro cameras were set to capture at full HD (1920x1080) resolution and at the widest angle available. They were placed on the harnesses rotated 90 degrees so as to capture the face of the other player as well as the blocks on the table (Fig.~\ref{fig:data_capture}(c)). 

In each session, the subjects play the game by standing at either end of a small rectangular table as shown in Fig.~\ref{fig:data_capture}(c). The person supervising the data collection enters player information and other meta-data about the game session into a form and then starts recording. He/she then instructs the players to begin their game session. They first manually activate the GoPro cameras to start recording and then clap their hands before starting their sessions. These claps were used to automatically synchronize the GoPro videos with the Kinect videos. The final dataset consists of the following data types for each game session:
\begin{enumerate}[itemsep=-1pt,topsep=-2pt, partopsep=1pt]
\item	Two Kinect videos (RGB)
\item	Two depth videos (depth encoded in RGB)
\item	Two GoPro videos (distortion corrected)
\item	Intrinsic and extrinsic calibrations for the two Kinect cameras
\item	Intrinsic calibrations and video frame aligned sequences of camera poses for the GoPro cameras
\item	Kinect tracked skeletons for the two participants
\item	Face and head pose tracking for the two individuals from the GoPro cameras when visible
\item	Eye Gaze information (3d vectors) for the two participants whenever available
\item	Object positions (2d bounding boxes, not 3d positions) and tracks for all the blocks within each gaming session. \\
\end{enumerate}
\noindent{\bf Data Annotation:} Since our focus is on \textit{joint attention} and \textit{entrainment}, we annotated 112 videos which were divided into 1213 10-second segments indicating the presence or absence of these two behaviors in each segment. To annotate the videos, we developed an innovative annotation schema drawn from concepts in the social psychology literature \cite{Adamson_JADD2012, Bernieri_JNVB1988}. The annotation schema is a series of questions, that could be used as a guideline to assist the annotators. The annotation schema associates high level social interaction predicates with more objectively perceptible measures. For example, \textit{Joint attention} is the shared focus of two individuals on a common subject and it involves eye gaze (on a person and on an object) and body language. Similarly, \textit{entrainment} is the alignment in the behavior of two individuals and it involves simultaneous movement, tempo similarity, and coordination. Each measure was rated using a low, medium, high measure for the entire 10 second segment. We hired six undergraduate sociology and psychology students to annotate the videos. The students were given a general introduction to the survey instrument and were then asked to code representative samples of the videos. The videos were annotated after ensuring that all the students as a group were annotating the sample videos accurately and reliably.

The dataset will be released with the acceptance of this paper.  We will also publish a fully detailed description of the collection, capture, and annotation. 
\begin{figure}
\centering
\begin{subfigure}[h]{0.45\textwidth}
\centering
\includegraphics[width=\textwidth]{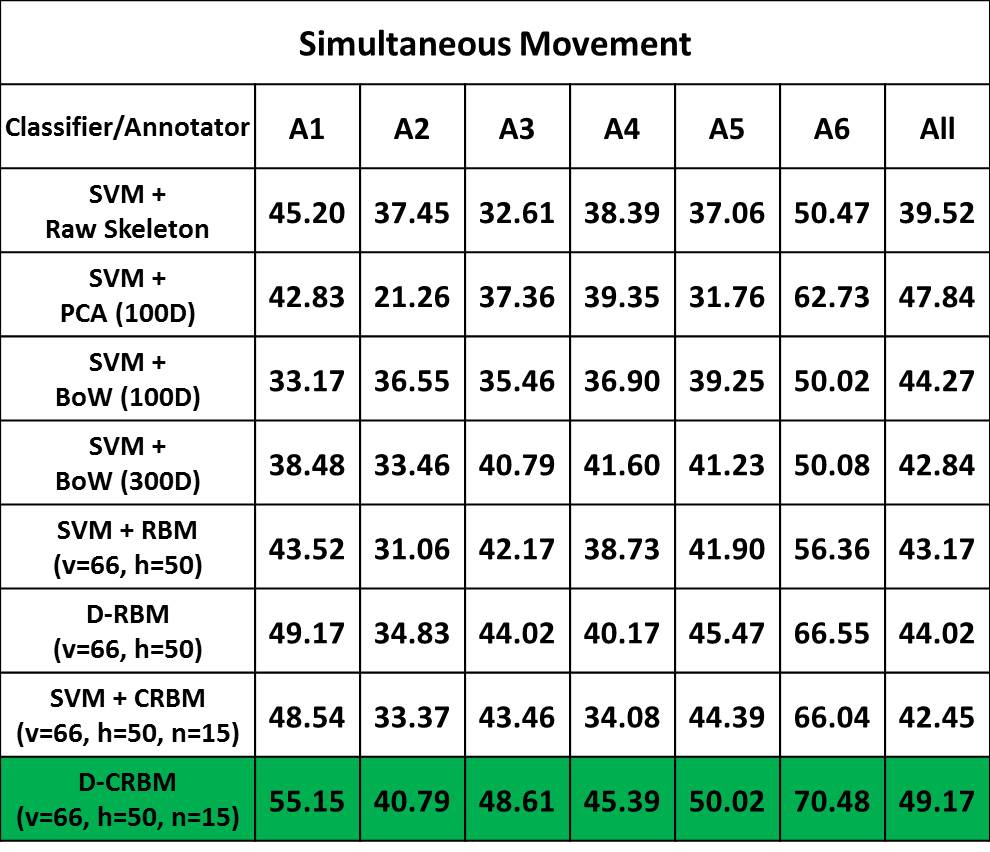}
\caption{Simultaneous Movement}
\label{fig:SimMovement}
\end{subfigure}
\vfill%
\begin{subfigure}[h]{0.45\textwidth}
\centering
\includegraphics[width=\textwidth]{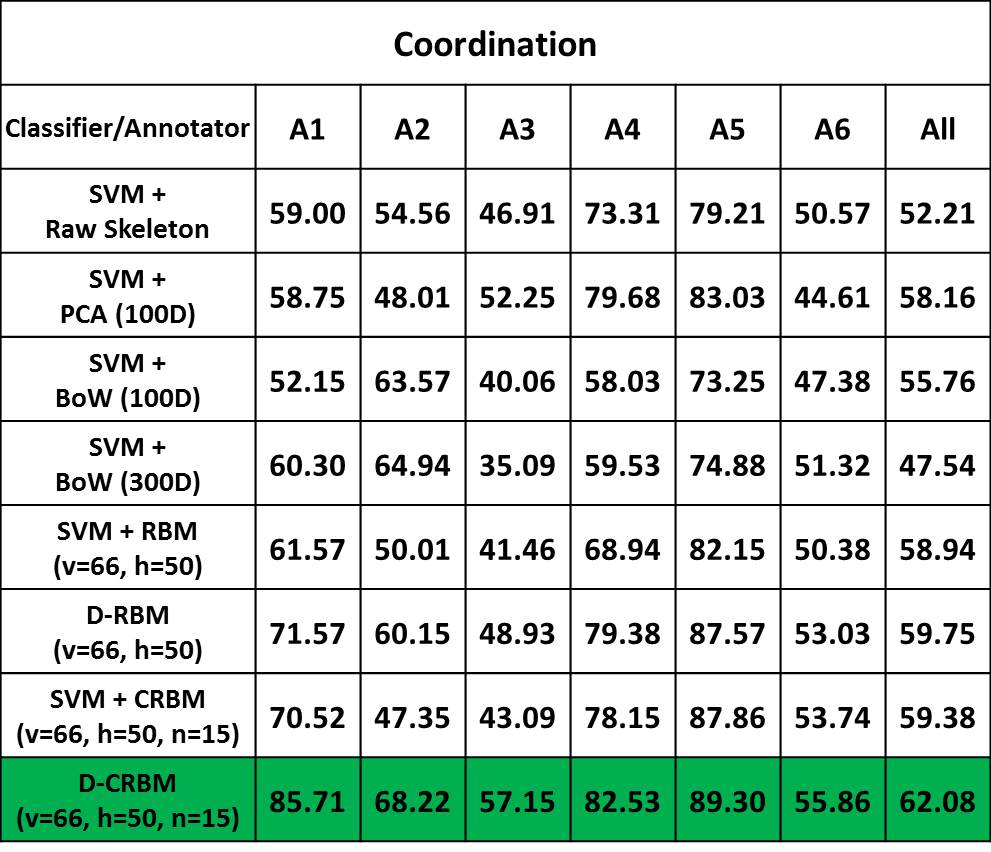}
\caption{Coordination}
\label{fig:Coordination}
\end{subfigure}
\vfill%
\begin{subfigure}[h]{0.45\textwidth}
\centering
\includegraphics[width=\textwidth]{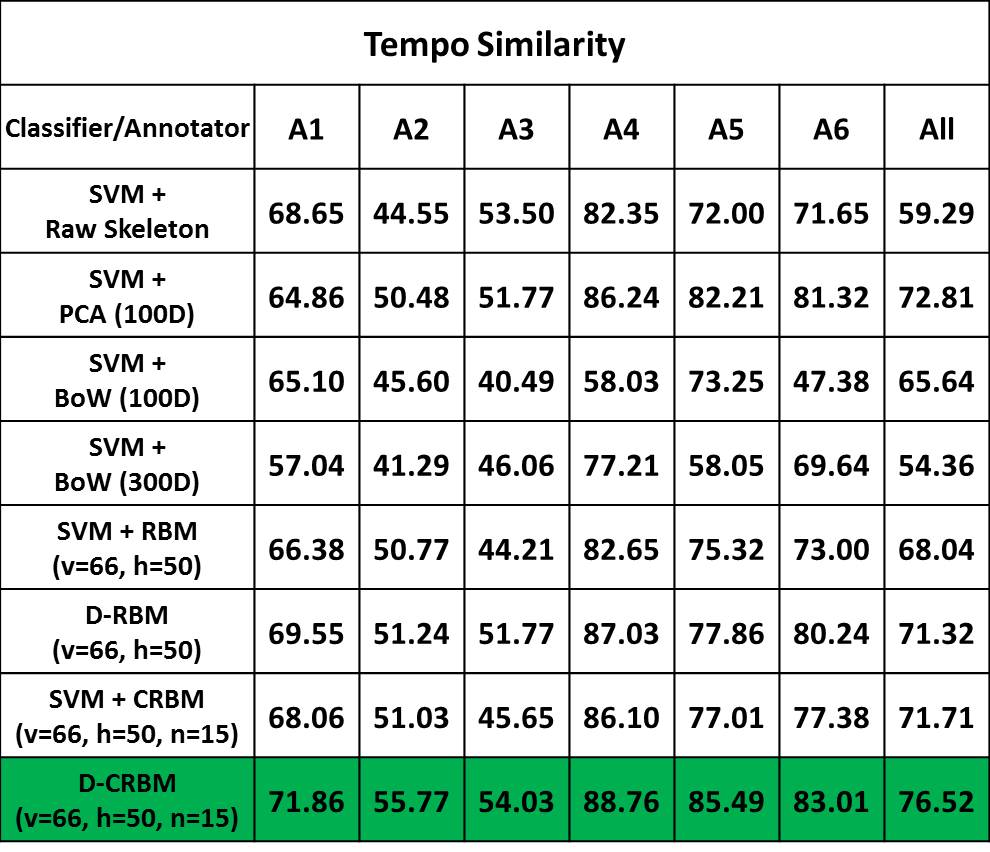}
\caption{Tempo Similarity}
\label{fig:TempoSim}
\end{subfigure}
\caption{Average classification results on ESIPs. It is clear that the DCRBM outperforms all other baselines on the three ESIPs.}
\label{fig:DCRBMClassification}
\end{figure}
\subsection{Quantitative Results}
In this section we describe the set of experiments we conducted to evaluate our proposed model.\\
\\
\noindent{\bf Implementation Details:} For our experiments, we relied only on the skeleton features. We use the 11 joints from the upper body of the two players since the tower game almost entirely involves only upper body actions. 

Using the 11 joints we extracted a set of first order static and dynamic handcrafted skeleton features. The static features are computed per frame. The features consist of, relationships between all pairs of joints of a single actor, as well as the relationships between all pairs of joints of both the actors. The dynamic features are extracted per window (a set of 300 frames). In each window, we compute first and second order dynamics (velocities and accelerations) of each joint, as well as relative velocities and accelerations of pairs of joints per actor, and across actors.  The dimensionality of the static and dynamic features is (257400 D).  To reduce their dimensionality we use Principle Component Analysis (PCA) (100 D), Bag-of-Words (BoW) (100 and 300 D) \cite{NieblesWF08}.  We also extracted Deep Learning features using RBMs and CRBMs (50 dimensions) 

For the DRBM and DCRBM we used the raw joint locations normalized with respect to a selected origin point.  We used the same dimensionality for both models $D(v)=66, D(h)=50$. For DCRBM we empirically evaluated history windows of different sizes, and found that a window of size $n=15$ works the best.\\
\begin{figure}
\centering
\begin{subfigure}{0.48\textwidth}
\centering
\includegraphics[width=0.7\textwidth]{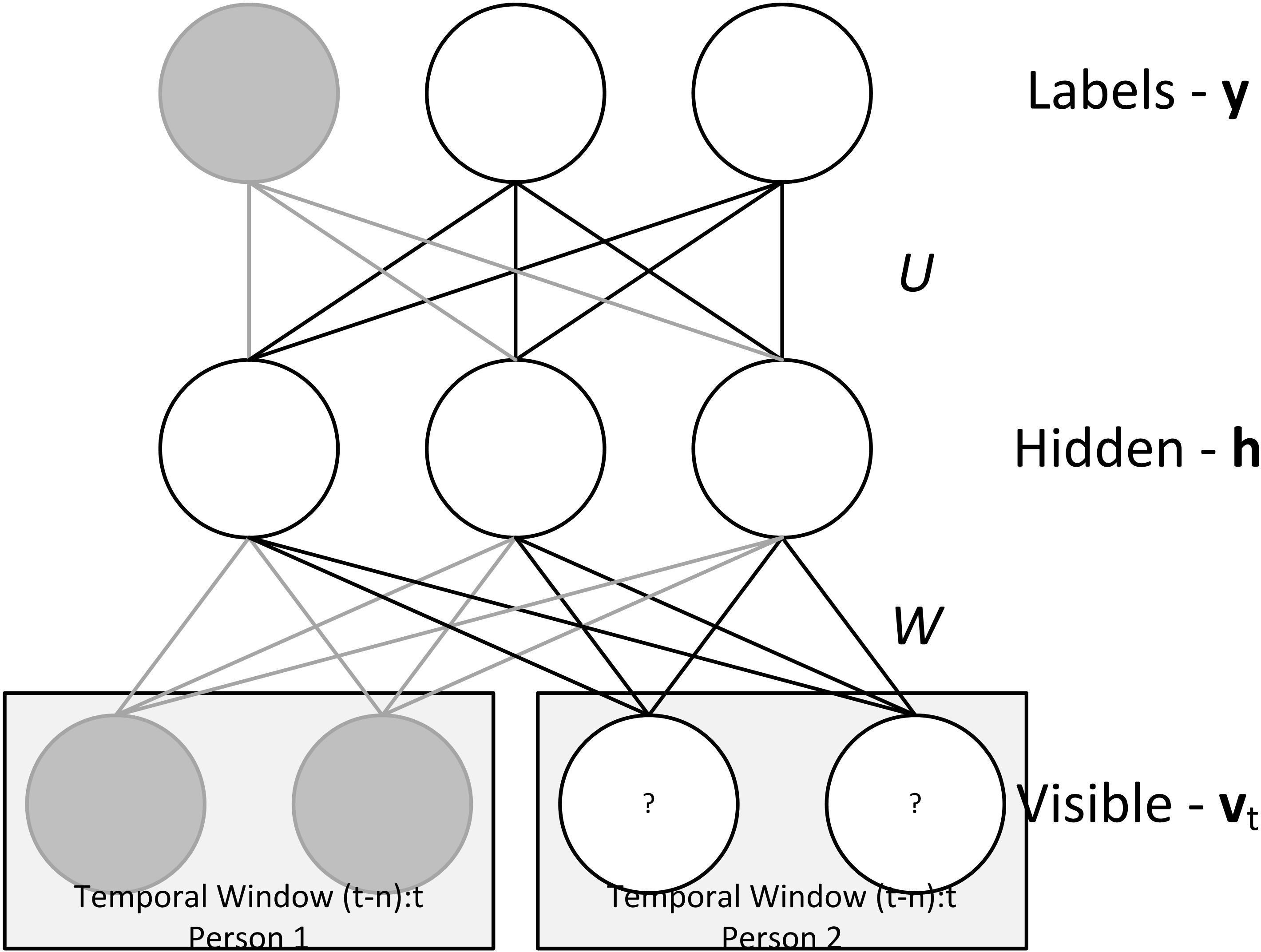}
\caption{Generating partial visible layer data }
\label{fig:CrossModal1}
\end{subfigure}
\vfill%
\begin{subfigure}{0.48\textwidth}
\centering
\fbox{\includegraphics[width=\textwidth]{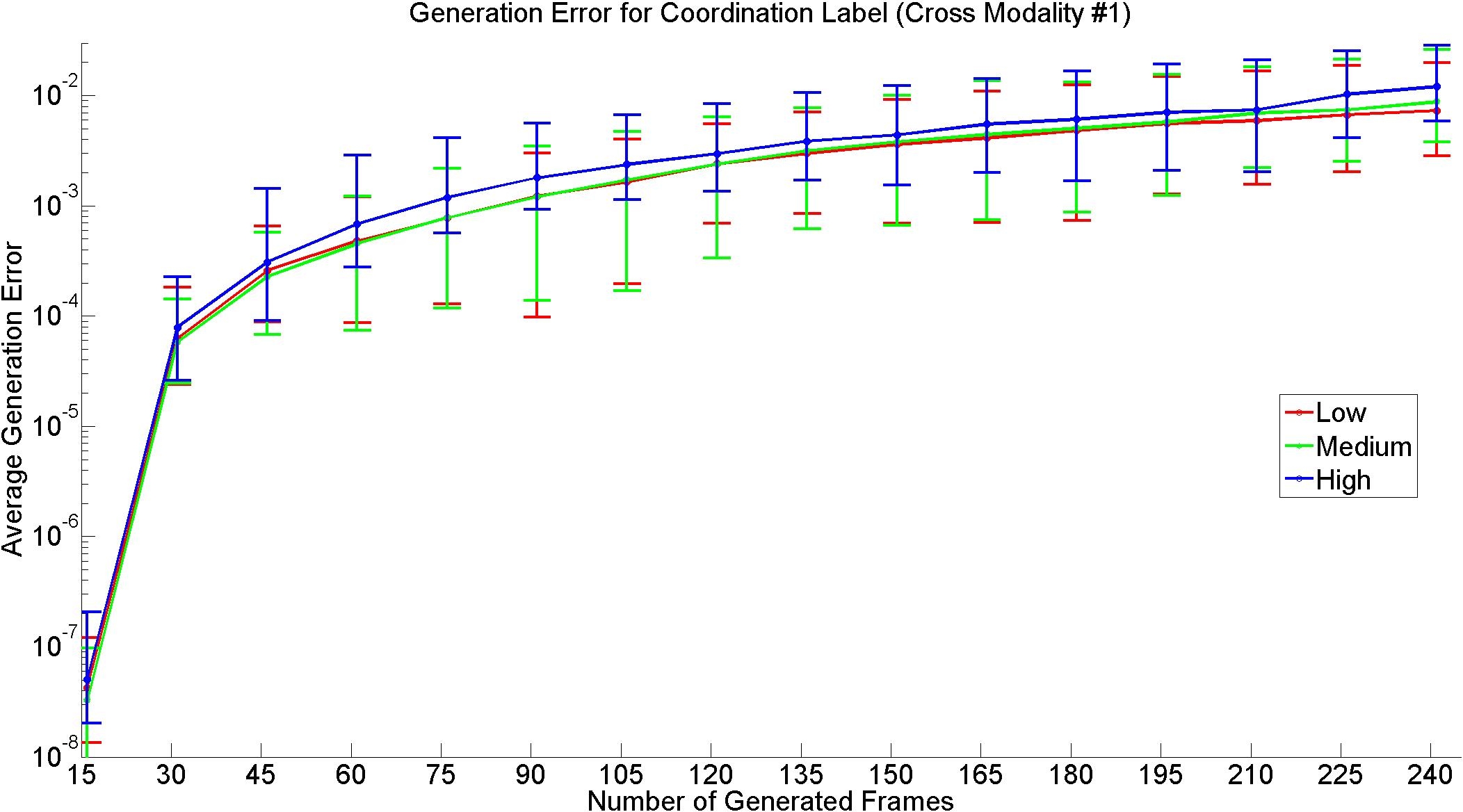}}\vfill%
\fbox{\includegraphics[width=\textwidth]{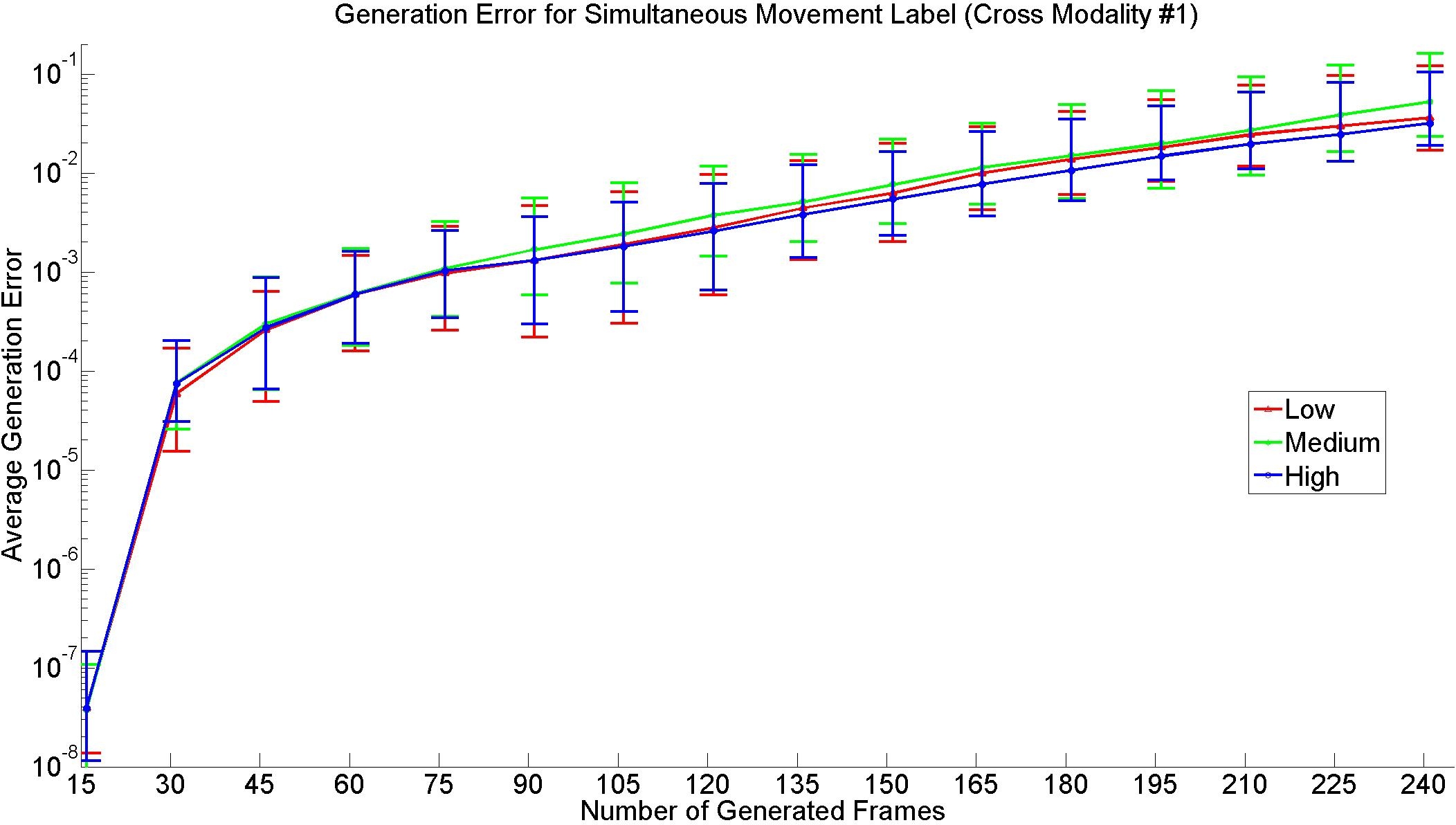}}\vfill%
\fbox{\includegraphics[width=\textwidth]{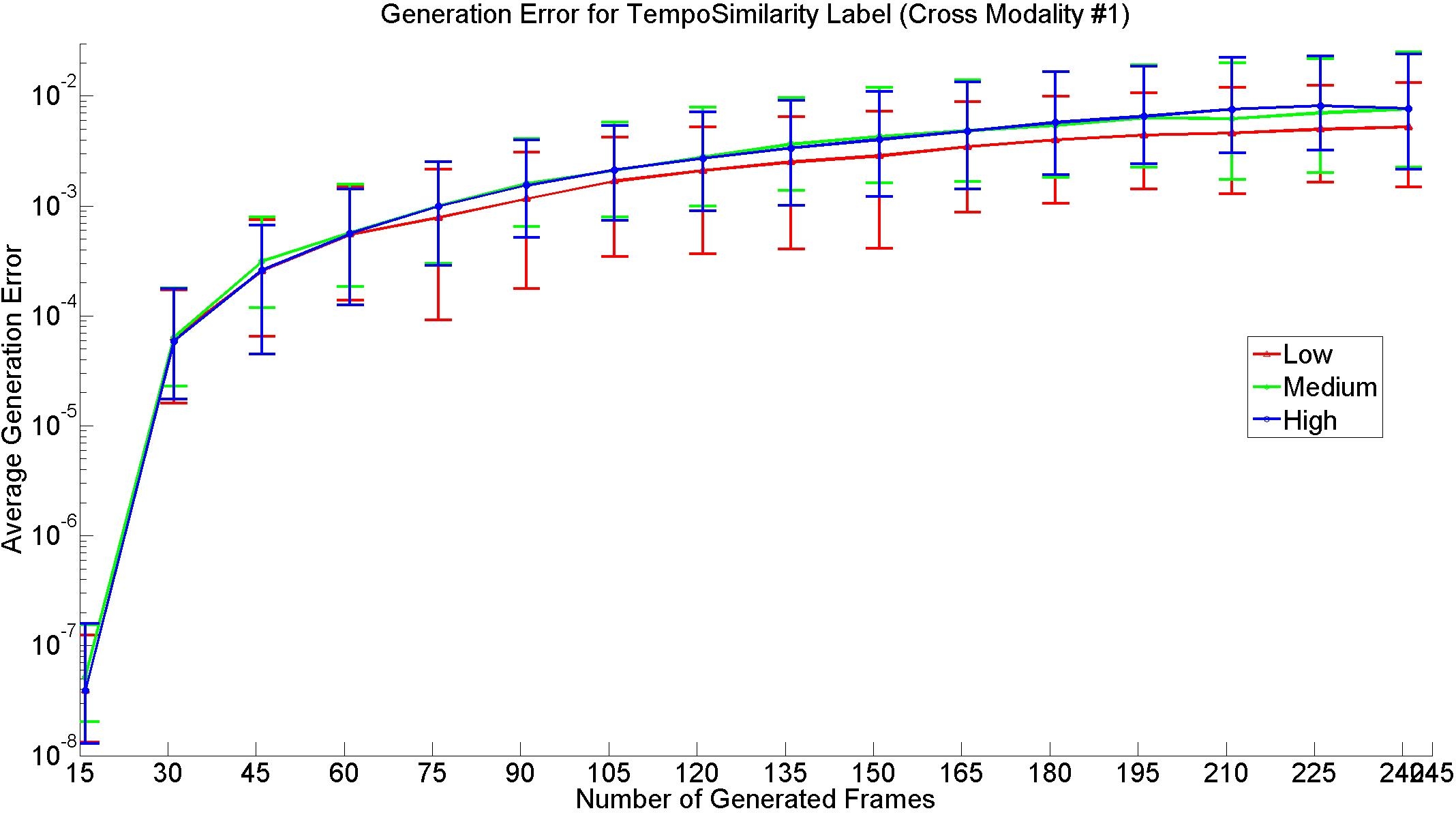}}
\caption{Average generation error (missing partial visible)}
\label{fig:CrossModal2}
\end{subfigure}
\caption{Average generation error for the partial visible layer by varying the generated window size for the three different ESIPs.}
\label{fig:DCRBMGenerationCross}
\end{figure}

\begin{figure}
\centering
\begin{subfigure}{0.48\textwidth}
\centering
\includegraphics[width=0.7\textwidth]{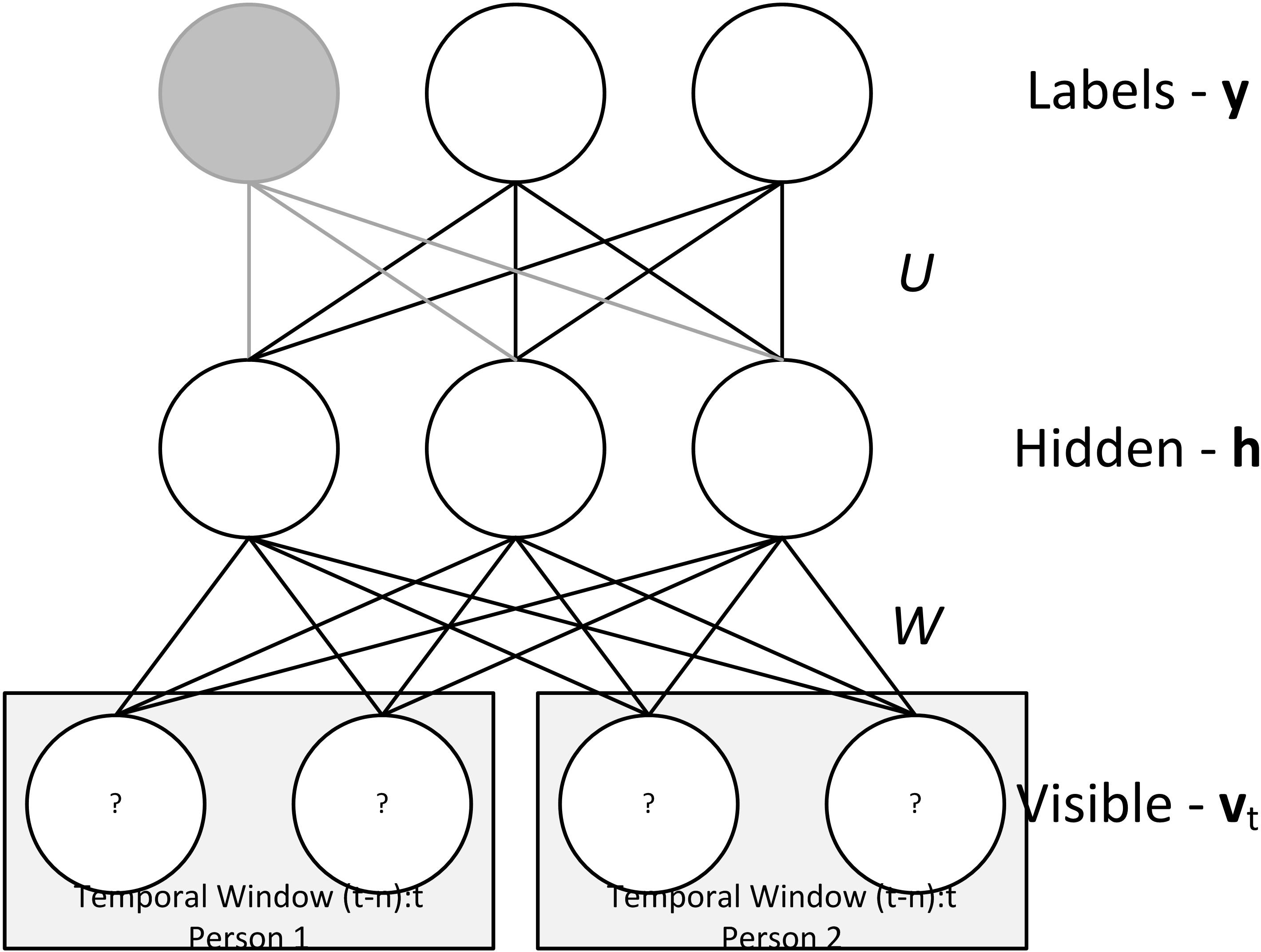}
\caption{Generating full visible layer data}
\label{fig:UniModal1}
\end{subfigure}
\vfill%
\begin{subfigure}{0.48\textwidth}
\centering
\fbox{\includegraphics[width=\textwidth]{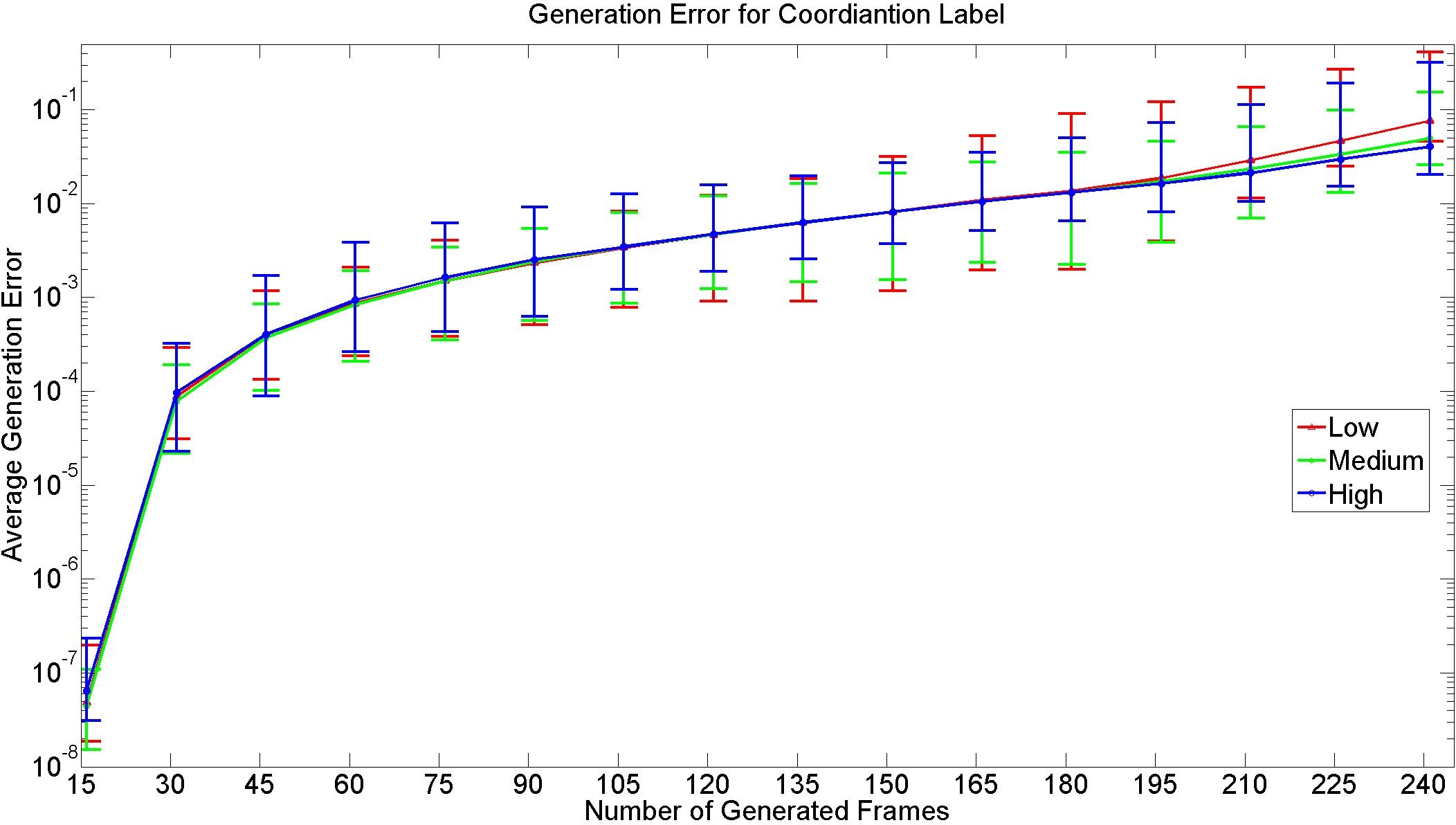}}\vfill%
\fbox{\includegraphics[width=\textwidth]{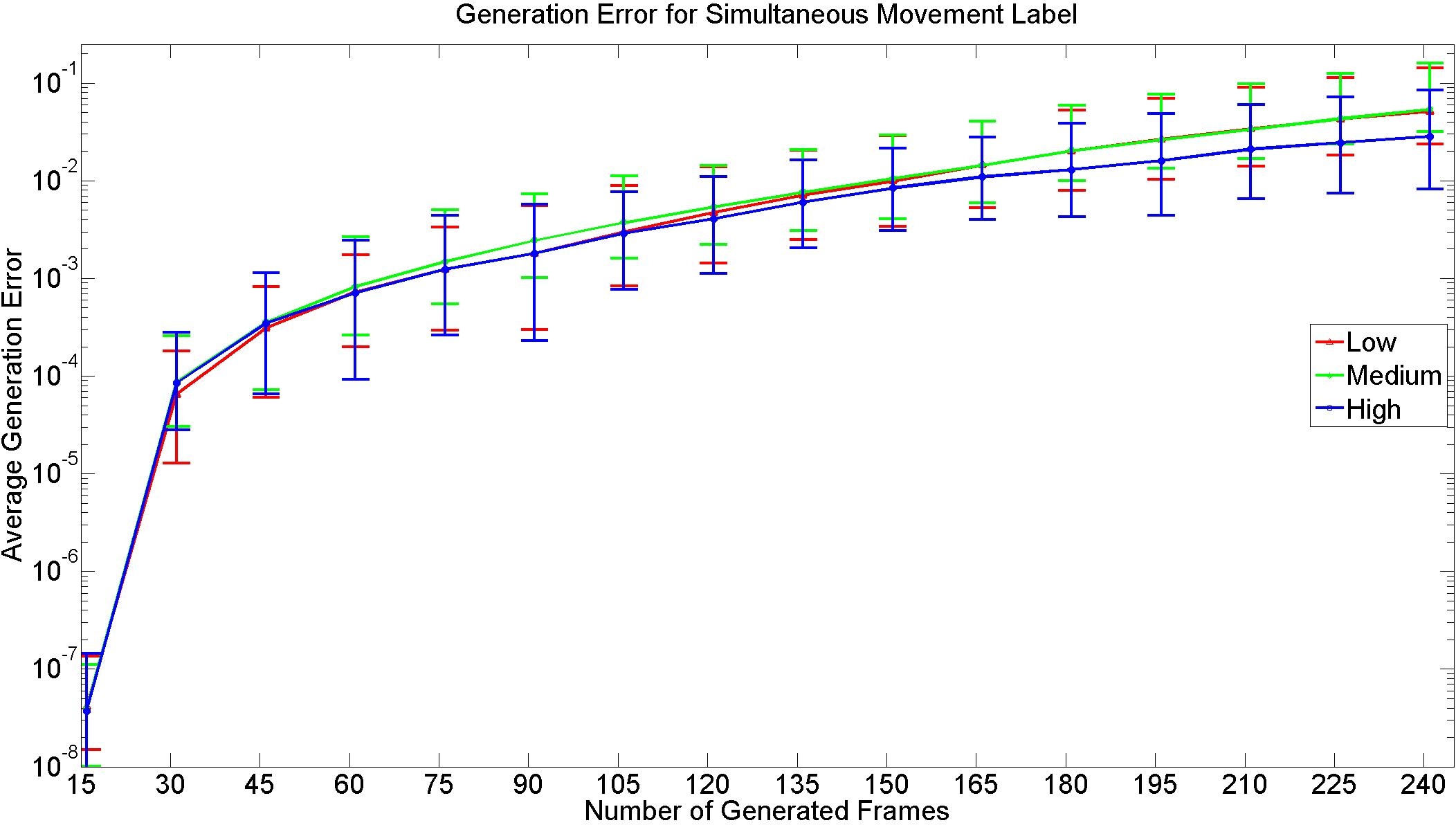}}\vfill%
\fbox{\includegraphics[width=\textwidth]{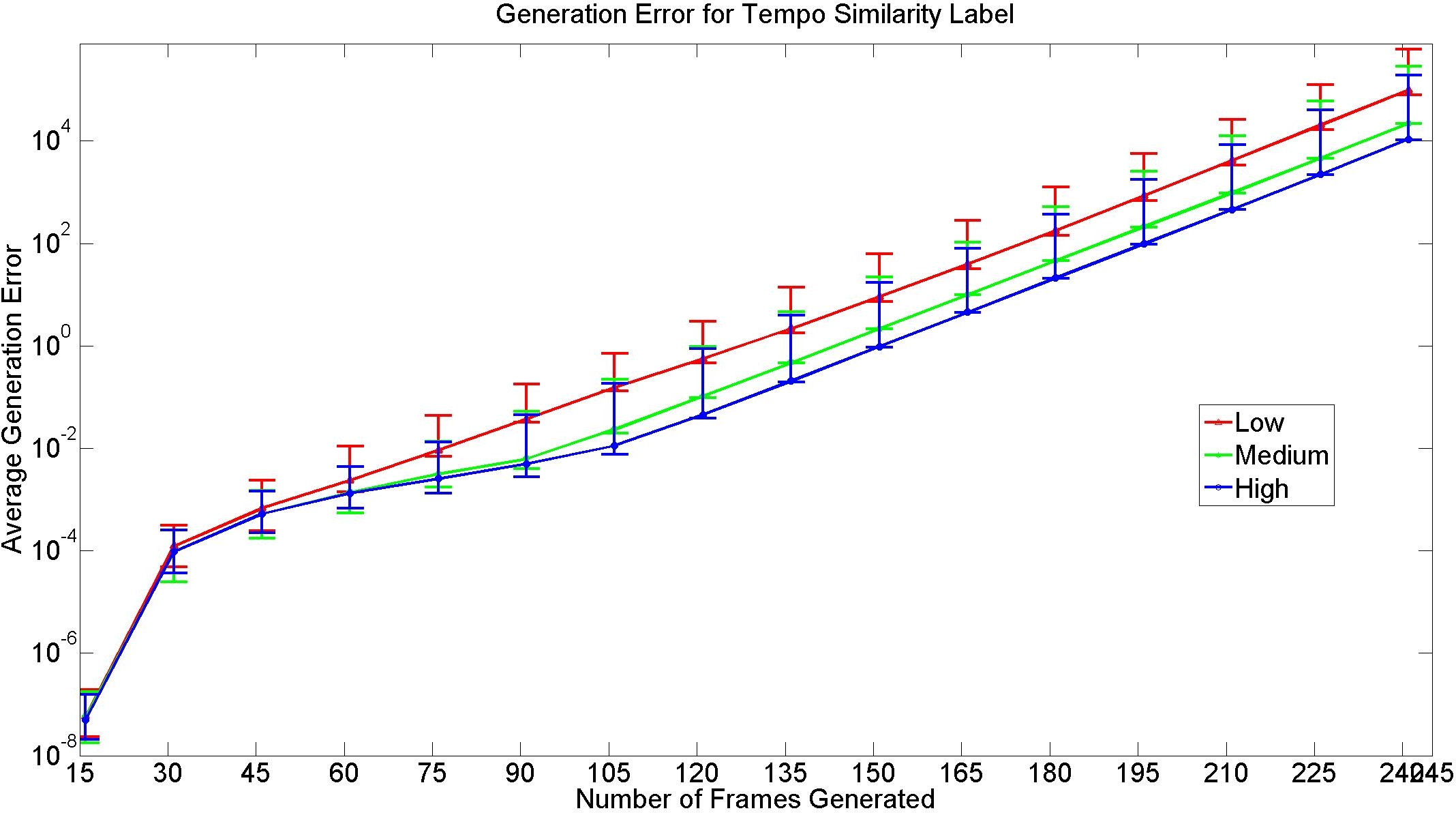}}
\caption{Average generation error (missing full visible)}
\label{fig:UniModal2}
\end{subfigure}
\caption{Average generation error for the full visible layer by varying the generated window size for the three different ESIPs.}
\label{fig:DCRBMGenerationUni}
\end{figure}
\noindent{\bf Results:} For the purpose of this paper we focused on the three ECIPs, Coordination, Simultaneous Movement, and Tempo Similarity. As a baseline we used a multi-class Support Vector Machine and the different types of features defined above to classify a certain ECIP.

We divided our evaluation into two tasks. The first task is the \textit{Classification Task}. We use the raw features of the two players and our goal is to predict the level (strength) of the three ECIPs. Each ECIP can be \textit{low}, \textit{medium} or \textit{high}, hence random classification accuracy is 33\%. The data is split into two sets, a training set consisting of 70\% of the instances, and a test set consisting of the remaining 30\%. We performed a 5 fold cross validation to guarantee unbiased results. Figure~\ref{fig:DCRBMClassification} shows our average classification accuracy on the Tower Game Dataset using different features and baselines combinations as well as the results from our DCRBM model. The evaluation is done with respect to the six annotators $\{A_1, A_2, \ldots , A_6\}$ as well as the mean annotation. We can see that the DCRBM model outperforms all the other models for each of the three measures across all annotators, thereby demonstrating its effectiveness on detecting these entrainment measures. Furthermore, the DCRBM model outperforms the PCA and BoW based features which are derived from the high dimensional handcrafted features, demonstrating its ability to learn a rich representation starting from the raw skeleton features. Finally, the performance of the DCRBM model indicates that the joint learning and inference of DCRBMs is superior to the staged approach of the SVM + CRBM model.   

The second task is the \textit{Generation Task}, where we are given the class label and our goal is to generate the data (i.e. the raw features) for that label. This task allows us to visualize what the classifier has learned. For generation, we initialize the model using 15 frames for each person, and then generate sequences of lengths varying from 16 to 300 frames. We measure the mean error between the ground-truth data and the generated data for each class label over 50 video instances. For this experiment, we evaluated generated sequences of varying length using a normalized mean squared error metric defined in (\ref{eqn:GenError}).
\begin{equation}
\text{Generation Error}=\left(\frac{\|\bf{v}_\text{Generated}-\bf{v}_\text{Groundtruth}\|}{\|\bf{v}_\text{Groundtruth}\|}\right)^2
\label{eqn:GenError}
\end{equation}
Generation is done in two different settings. In the first setting, given partial visible player data (one player's features) as well as the class label, the goal is to generate the other player's data. Figure~\ref{fig:DCRBMGenerationCross} shows our average generation error using our DCRBM model for generating the partial visible layer.  In the second setting, given only the class label, the goal is to generate the entire visible layer data (i.e. the raw features for both the players). Figure~\ref{fig:DCRBMGenerationUni} shows our average generation error on using our DCRBM model for generating the full visible layer. We can see that the generation is relatively low $(< 0.1)$ in all cases (except for Tempo Similarity\footnote{Tempo Similarity measures the similarity in the rate of the motion of the two players, and when data from both the players is missing generating their raw features based on whether their rate of motion is similar is extremely under constrained} when generating the entire visible layer data) demonstrating the effectiveness of DCRBM model for generating data. Also, the error is similar across different levels (strengths) for each measure indicating that the model is relatively stable. Finally, the error increases with the length of the generated sequence, which is expected as the possibility of variation in the ground-truth sequences increases with length.  

Therefore, the classification task shows that DCRBMs can effectively detect the constituents of \textit{entrainment} (an ESIP). Similarly, the generation task shows that DCRBMs can effectively generate raw skeleton data of the actors while modeling the different strengths of each constituent (measure).  

\section{Conclusions and Future Work}
\label{sec:conclusion}
We presented a novel approach to computational modeling of social interactions based on modeling of essential social interaction predicates (ESIPs) such as joint attention and entrainment. Our data collection was guided by social psychological theory and methodology. We introduce a new ``Tower Game" dataset consisting of audio-visual capture of dyadic interactions labeled with the ESIPs, that should spark new research in computational social interaction modeling. We proposed a novel joint Discriminative Conditional Restricted Boltzmann Machine (DCRBM) model that enabled us to uncover actionable constituents of the ESIPs in two steps. First, we trained the DCRBM model and second, used it to generate lower-level data corresponding to ESIP's with high accuracy. 

Such purely computational decomposition of ESIPs into actionable behavioral constituents is unprecedented and powerful, and offers rich possibilities for further research. First, we can substantially advance the understanding of ESIPs by uncovering mid-level predicates using the hidden layers of the DCRBM thus going beyond the current low-level feature generation to a multi-level understanding of the semantics of ESIPs. Second, we would like to extend our framework to multimodal streams that also include gaze, facial behaviors, head pose and audio so as to get a full understanding of actionable behaviors that make up the ESIPs. For instance, we may find out that coordinating gaze and gestural behavior is the most effective in establishing rapport, or perhaps not. Third, such a comprehensive multimodal and semantic model would capture the overall ``rules of engagement" in a social interaction. Such a model would therefore lend itself to monitoring and training applications such as automatic assessment of the efficacy of an interaction in terms of establishment of rapport-engagement and generation of ``interaction-realistic'' avatar behaviors in a virtual reality environment that convey realism in terms of interaction dynamics rather than through photo or audio realism, and thus achieve immersion and engagement, as well as more efficacious human-robot interaction. We have thus laid the foundation of a computational approach that enables us to move from ``folklore" based methods of establishing ESIPs to methods that are systematically arrived at through computational analysis of data from scientific observations.

\section*{Acknowledgments}
This work is supported by DARPA W911NF-12-C-0001. The views,  opinions,  and/or conclusions contained in this paper are those of the author and should not be interpreted as  representing  the  official  views  or  policies,  either  expressed or implied of the DARPA or the DoD.

\bibliographystyle{abbrv}
\bibliography{References}  % sigproc.bib is the name of the Bibliography in this case

\begin{thebibliography}{10}

\bibitem{Adamson_JADD2012}
L.~Adamson and et~al.
\newblock Rating parent-child interactions: Joint engagement, communication
  dynamics and shared topics in autism, down syndrome, and typical development.
\newblock {\em JADD}, 2012.

\bibitem{Amer_WACV2014}
M.~Amer, B.~Siddiquie, S.~Khan, A.~Divakaran, and H.~Sawhney.
\newblock Multimodal fusion using dynamic hybrid models.
\newblock In {\em WACV}, 2014.

\bibitem{Amer_ICASSP2014}
M.~Amer, B.~Siddiquie, C.~Richey, and A.~Divakaran.
\newblock Emotion detection in speech using deep networks.
\newblock In {\em ICASSP}, 2014.

\bibitem{Baron_MIT1997}
S.~Baron-Cohen.
\newblock Mindblindness: An essay on autism and theory of mind.
\newblock In {\em MIT}, 1997.

\bibitem{Bengio_FTML2009}
Y.~Bengio.
\newblock Learning deep architectures for ai.
\newblock In {\em FTML}, 2009.

\bibitem{Bernieri_JNVB1988}
F.~J. Bernieri.
\newblock Coordinated movement and rapport in teacher-student interactions.
\newblock {\em Journal of Non-Verbal Behavior}, 1988.

\bibitem{Bosch_TPAMI2008}
A.~Bosch, A.~Zisserman, and M.~Xavier.
\newblock Scene classification using a hybrid generative/discriminative
  approach.
\newblock In {\em TPAMI}, 2008.

\bibitem{Calvo_TAC2010}
R.~Calvo and S.~D'Mello.
\newblock Affect detection: An interdisciplinary review of models, methods, and
  their applications.
\newblock In {\em IEEE Transactions on Affective Computing}, 2010.

\bibitem{Chaquet_CVIU2013}
J.~M. Chaquet, E.~J. Carmona, and A.~Fern{\'a}ndez-Caballero.
\newblock A survey of video datasets for human action and activity recognition.
\newblock {\em CVIU}, 117(6):633 -- 659, 2013.

\bibitem{Jaegher_TCS2010}
H.~De~Jaegher, E.~Di~Paolo, and S.~Gallagher.
\newblock Can social interaction constitute social cognition?
\newblock {\em Trends in Cognitive Science}, 2010.

\bibitem{DMello_IS2007}
S.~D'Mello, R.~W. Picard, and A.~Graesser.
\newblock Toward an affect-sensitive autotutor.
\newblock In {\em IEEE Intelligent Systems}, 2007.

\bibitem{Druck_ICML2010}
G.~Druck and A.~McCallum.
\newblock High-performance semi-supervised learning using discriminatively
  constrained generative models.
\newblock In {\em ICML}, 2010.

\bibitem{Fanelli_IJCV2012}
G.~Fanelli, M.~Dantone, J.~Gall, A.~Fossati, and L.~V. Gool.
\newblock Random forests for real time 3d face analysis.
\newblock In {\em IJCV}, 2012.

\bibitem{Fantasia_FP2014}
V.~Fantasia, H.~D. Jaegher, and A.~Fasulo.
\newblock We can work it out: an enactive look at cooperation.
\newblock {\em Frontieres in Psychology}, 2014.

\bibitem{Fujino_PAMI2008}
A.~Fujino, N.~Ueda, and K.~Saito.
\newblock Semi-supervised learning for a hybrid generative/discriminative
  classifier based on the maximum entropy principle.
\newblock In {\em TPAMI}, 2008.

\bibitem{Garg_ACL2011}
N.~Garg and J.~Henderson.
\newblock Temporal restricted boltzmann machines for dependency parsing.
\newblock In {\em ACL}, 2011.

\bibitem{Ghosh_ICMI2014}
S.~Ghosh, M.~Chatterjee, and L.-P. Morency.
\newblock A multimodal context-based approach for distress assessment.
\newblock In {\em ICMI}, 2014.

\bibitem{Hausler_CoRR2012}
C.~Hausler and A.~Susemihl.
\newblock Temporal autoencoding restricted boltzmann machine.
\newblock In {\em CoRR}, 2012.

\bibitem{Hinton_NC2002}
G.~E. Hinton.
\newblock Training products of experts by minimizing contrastive divergence.
\newblock In {\em NC}, 2002.

\bibitem{Hinton_NC2006}
G.~E. Hinton, S.~Osindero, and Y.~W. Teh.
\newblock A fast learning algorithm for deep belief nets.
\newblock In {\em NC}, 2006.

\bibitem{Jebara_MLR2004}
T.~Jebara and et. al.
\newblock Probability product kernels.
\newblock In {\em MLR}, 2004.

\bibitem{Koppula_ICML2013}
H.~S. Koppula and A.~Saxena.
\newblock Learning spatio-temporal structure from rgb-d videos for human
  activity detection and anticipation.
\newblock In {\em ICML}, 2013.

\bibitem{Lan_PAMI2012}
T.~Lan, Y.~Wang, W.~Yang, S.~Robinovitch, and G.~Mori.
\newblock Discriminative latent models for recognizing contextual group
  activities.
\newblock In {\em PAMI}, 2012.

\bibitem{Larochelle_ICML2008}
H.~Larochelle and Y.~Bengio.
\newblock Classification using discriminative restricted boltzmann machines.
\newblock In {\em ICML}, 2008.

\bibitem{Lasserre_CVPR2006}
J.~Lasserre, C.~Bishop, and T.~Minka.
\newblock Principled hybrids of generative and discriminative models.
\newblock In {\em CVPR}, 2006.

\bibitem{Levinson_FARS2006}
S.~C. Levinson.
\newblock On the human "interaction engine".
\newblock In {\em Roots of Human Sociality Culture, Cognition and Interaction}.
  Berg, 2006.

\bibitem{Lewandowski_ICML2009}
N.~B. Lewandowski, Y.~Bengio, and P.~Vincent.
\newblock Modeling temporal dependencies in high-dimensional sequences:
  Application to polyphonic music generation and transcription.
\newblock In {\em ICML}, 2012.

\bibitem{Li_CVPR2011}
X.~Li, T.~Lee, and Y.~Liu.
\newblock Hybrid generative-discriminative classification using posterior
  divergence.
\newblock In {\em CVPR}, 2011.

\bibitem{Tian_PAMI2001}
Y.~li~Tian, T.~Kanade, and J.~F. Cohn.
\newblock Recognizing action units for facial expression analysis.
\newblock In {\em IEEE PAMI}. 2001.

\bibitem{Louwerse_CS2012}
M.~M. Louwerse, R.~Dale, E.~G. Bard, and P.~Jeuniaux.
\newblock Behavior matching in multimodal communication is synchronized.
\newblock {\em Cognitive Science}, 2012.

\bibitem{Zeiler_NIPS2011}
L.~S. M.~D.~Zeiler, G. W.~Taylor, I.~Matthews, and R.~Fergus.
\newblock Facial expression transfer with input-output temporal restricted
  boltzmann machines.
\newblock In {\em NIPS}, 2011.

\bibitem{Mccallum_AAAI2006}
A.~Mccallum, C.~Pal, G.~Druck, and X.~Wang.
\newblock Multi-conditional learning: Generative/discriminative training for
  clustering and classification.
\newblock In {\em AAAI}, 2006.

\bibitem{Memisevic_CVPR2007}
R.~Memisevic and G.~E. Hinton.
\newblock Unsupervised learning of image transformations.
\newblock In {\em CVPR}, 2007.

\bibitem{Mohamed_ICML2009}
A.~R. Mohamed and G.~E. Hinton.
\newblock Phone recognition using restricted boltzmann machines.
\newblock In {\em ICASSP}, 2009.

\bibitem{Mota_CVPRW2003}
S.~Mota and R.~W. Picard.
\newblock Automated posture analysis for detecting learner's interest level.
\newblock In {\em CVPRW}, 2003.

\bibitem{NieblesWF08}
J.~Niebles, H.~Wang, and L.~Fei-Fei.
\newblock Unsupervised learning of human action categories using
  spatial-temporal words.
\newblock {\em IJCV}, 79(3):299--318, 2008.

\bibitem{DiPaolo_FHN2012}
E.~D. Paolo and H.~D. Jaegher.
\newblock The interactive brain hypothesis.
\newblock {\em Frontiers in Human Neuroscience}, 2012.

\bibitem{Perina_PAMI2012}
A.~Perina and et~al.
\newblock Free energy score spaces: Using generative information in
  discriminative classifiers.
\newblock In {\em TPAMI}, 2012.

\bibitem{Picard_MIT1995}
R.~W. Picard.
\newblock {\em Affective Computing}.
\newblock MIT Press, 1995.

\bibitem{Ramirez_ACII2011}
G.~Ramirez, T.~Baltrusaitis, and L.~P. Morency.
\newblock Modeling latent discriminative dynamic of multi-dimensional affective
  signals.
\newblock In {\em ACII}, 2011.

\bibitem{Ranzato_CVPR2011}
M.~A. Ranzato and et. al.
\newblock On deep generative models with applications to recognition.
\newblock In {\em CVPR}, 2011.

\bibitem{Raptis_SCA2011}
M.~Raptis, D.~Kirovski, and H.~Hoppe.
\newblock Real-time classification of dance gestures from skeleton animation.
\newblock In {\em SCA}, 2011.

\bibitem{Rehg_CVPR2013}
J.~M. Rehg and et~al.
\newblock Decoding children's social behavior.
\newblock {\em CVPR}, 2013.

\bibitem{Ryoo_IJCV2009}
M.~Ryoo and J.~Aggarwal.
\newblock Semantic representation and recognition of continued and recursive
  human activities.
\newblock In {\em IJCV}, 2009.

\bibitem{Salakhutdinov_Science2006}
R.~Salakhutdinov and G.~E. Hinton.
\newblock Reducing the dimensionality of data with neural networks.
\newblock In {\em Science}, 2006.

\bibitem{Sebanz_TCS2009}
N.~Sebanz and G.~Knoblich.
\newblock Prediction in joint action: What, when, and where.
\newblock {\em Topics in Cognitive Science}, 2009.

\bibitem{Shotton_CVPR2011}
J.~Shotton and et~al.
\newblock Real-time human pose recognition in parts from a single depth image.
\newblock In {\em CVPR}, 2011.

\bibitem{Siddiquie_ICME2013}
B.~Siddiquie, S.~Khan, A.~Divakaran, and H.~Sawhney.
\newblock Affect analysis in natural human interactions using joint hidden
  conditional random fields.
\newblock In {\em ICME}, 2013.

\bibitem{Sminchisescu_CVPR2006}
C.~Sminchisescu, A.~Kanaujia, and D.~Metaxas.
\newblock Learning joint top-down and bottom-up processes for 3d visual
  inference.
\newblock In {\em CVPR}, 2006.

\bibitem{Sun_ACII2011}
X.~Sun, J.~Lichtenauer, M.~F. Valstar, A.~Nijholt, and M.~Pantic.
\newblock A multimod al database for mimicry analysis.
\newblock In {\em ACII}, 2011.

\bibitem{Sutskever_NIPS2008}
I.~Sutskever, G.~Hinton, and G.~Taylor.
\newblock The recurrent temporal restricted boltzmann machine.
\newblock In {\em NIPS}, 2008.

\bibitem{Sutskever_AISTATS2007}
I.~Sutskever and G.~E. Hinton.
\newblock Learning multilevel distributed representations for high-dimensional
  sequences.
\newblock In {\em AISTATS}, 2007.

\bibitem{Taylor_CVPR2010}
G.~W. Taylor and et. al.
\newblock Dynamical binary latent variable models for 3d human pose tracking.
\newblock In {\em CVPR}, 2010.

\bibitem{Taylor_JMLR2011}
G.~W. Taylor, G.~E. Hinton, and S.~T. Roweis.
\newblock Two distributed-state models for generating high-dimensional time
  series.
\newblock In {\em Journal of Machine Learning Research}, 2011.

\bibitem{Tomasello_Harvard2001}
M.~Tomasello.
\newblock {\em The Cultural Origins of Human Cognition}.
\newblock Harvard University Press, 2001.

\end{thebibliography}

\end{document}